\pdfoutput=1
\documentclass[11pt]{article}
\usepackage{jheppub}
\usepackage{amssymb}
\usepackage{amsmath}
\usepackage{amsfonts}
\usepackage{wasysym}

\allowdisplaybreaks

\usepackage[usenames,dvipsnames]{xcolor}
\usepackage{tikz}
\usetikzlibrary{patterns}
\usetikzlibrary{decorations.pathmorphing}
\usetikzlibrary{decorations.markings}
\usepgflibrary{arrows.meta}
\usetikzlibrary{intersections,through,hobby}
\usepackage{tkz-euclide}
\tikzset{
  WLBE/.style={double distance=1.1pt,postaction={decorate},
    decoration={markings,mark=at position .5 with {\arrow{Straight Barb[scale=0.5]}}}},  
  WLB/.style={double distance=1.1pt,postaction={decorate},
    decoration={markings,mark=at position .2 with {\arrow{Straight Barb[scale=0.5]}}}},
  WLBC/.style={double distance=1.1pt,postaction={decorate},
    decoration={markings,mark=at position .5 with {\arrow{Straight Barb[scale=0.5]}}}},
  WLBS/.style={double distance=1.1pt,postaction={decorate},
    decoration={markings,mark=at position .6 with {\arrow{Straight Barb[scale=0.5]}}}},
  GLUON/.style={decorate,
    decoration={coil, amplitude=2.5pt,segment length=3.25pt, aspect=0.65}},
  sglu/.style={
    draw=none,
    decoration={name=none},
    postaction={
      draw,
      line width = 0.6pt,
      decoration={
        coil,
        aspect=0.85,
        mirror,
        amplitude=2.0pt,
        segment length=4pt,
      },
      decorate=true},
    color=Gray},
  higgs/.style={line width=0.8pt,Gray, densely dashed},
  sghost/.style={line width=0.8pt,Gray, densely dotted,postaction={decorate},
    decoration={markings,mark=at position .4 with {\arrow{latex[scale=0.5]}}}},
  squark/.style={line width=0.8pt,Gray, postaction={decorate},
    decoration={markings,mark=at position .4 with {\arrow{latex[scale=0.5]}}}}
}

\definecolor{colA}{HTML}{c19277}
\definecolor{colB}{HTML}{e1bc91}
\definecolor{colD}{HTML}{62959c}
\hypersetup{
  citecolor=colA, 		%color of links to bibliography
  linkcolor=colA,    	%color of internal links
  urlcolor=colD,			%color of external links
}

\newcommand{\be}{\begin{equation}}
\newcommand{\ee}{\end{equation}}
\newcommand{\ep}{\ensuremath{\varepsilon}}
\newcommand{\dm}{\ensuremath{\textrm{d}}}

\newcommand{\pent}[2]{\ensuremath{ \mbox{\Large \pentagon}^{#1}_{#2}}}
\newcommand{\pbox}[2]{\ensuremath{ \mbox{\Large \Square}^{#1}_{#2}}}
\newcommand{\dFddA}{\ensuremath{\textrm{d}\Phi_{\delta\delta}^{nn}}}

\newcommand{\dFtdA}{\ensuremath{\textrm{d}\Phi_{\theta\delta}^{nn}}}
\newcommand{\dFttB}{\ensuremath{\textrm{d}\Phi_{\theta\theta}^{n\bar{n}}}}

\newcommand{\nf}{\ensuremath{n_f}}
\newcommand{\NC}{\ensuremath{N_c}}

\newcommand{\CF}{\ensuremath{C_F}}
\newcommand{\cA}{\ensuremath{C_A}}
\newcommand{\cR}{\ensuremath{C_R}}
\newcommand{\TF}{\ensuremath{T_F}}
\begin{document}

\title{One-loop corrections to the double-real emission contribution
  to the zero-jettiness soft function at N3LO in QCD
}

\preprint{
\begin{minipage}[c]{0.3\linewidth}
  \begin{flushright}
    MPP-2024-2\\
    P3H-24-003\\
    TTP24-001\\
    TUM-HEP-1493/24\\
    ZU-TH 03/24
  \end{flushright}
\end{minipage}}

\author[a]{Daniel Baranowski,}
\author[b]{Maximilian Delto,}
\author[c]{Kirill~Melnikov,}
\author[c]{Andrey~Pikelner}
\author[d]{and Chen-Yu Wang}

\affiliation[a]{Physik Institut, Universität Zürich, Winterthurerstrasse 190, 8057 Zürich, Switzerland}
\affiliation[b]{Physics Department, Technical University of Munich, James-Franck-Strasse 1,  85748, Munich, Germany}
\affiliation[c]{Institute for Theoretical Particle Physics (TTP), Karlsruhe Institute of
  Technology, 76128, Karlsruhe, Germany
}
\affiliation[d]{Max-Planck Institute for Physics, Boltzmannstr.\ 8, 85748 Garching, Germany}

\emailAdd{daniel.baranowski@physik.uzh.ch}
\emailAdd{maximilian.delto@tum.de}
\emailAdd{kirill.melnikov@kit.edu}
\emailAdd{andrey.pikelner@kit.edu}
\emailAdd{cywang@mpp.mpg.de}

\abstract{We present an analytic calculation of the one-loop correction to the
double-real emission contribution to the zero-jettiness soft function at N3LO in
QCD, accounting for both gluon-gluon and quark-antiquark soft final-state
partons. We explain all the relevant steps of the computation including the
reduction of phase-space integrals to master integrals in the presence of
Heaviside functions, and the methods we employed to compute them. }

\maketitle

\section{Introduction}

Description of hard scattering processes at the LHC relies on QCD perturbation
theory. To arrive at finite predictions for infra-red safe observables, one
needs to combine all contributions to a particular cross section, that are
proportional to the strong coupling constant raised to a certain power, but may
differ by the number of final-state partons. This step is non-trivial as
contributions with different partonic multiplicities suffer from infrared and
collinear divergencies that only cancel in the sum
\cite{Bloch:1937pw,Kinoshita:1962ur,Lee:1964is}. One of the several ways to
simplify this step is known as the slicing method where one chooses a kinematic
parameter $h$ that distinguishes contributions with Born kinematics from those
with additional final-state partons. Assuming that $h=0$ corresponds to Born
kinematics, one then computes the cross section for small values of $h$, where
the kinematics is very much Born-like, and only soft or collinear final-state
partons can be present in addition to the hard ones that already appear in the
Born process.

 Since description of soft and collinear emissions is independent of the hard
process, it is not surprising that cross sections at small values of the slicing
parameter are computed by integrating universal functions that describe soft-
and collinear limits of squared matrix elements over phase spaces available to
unresolved partons. For certain slicing variables, one can prove that these
integrals factorize into so-called soft, beam and jet functions and, once these
functions are known, singular dependencies of cross sections on $h$ can be
reconstructed. Furthermore, since soft, beam and jet functions separately
satisfy renormalization group equations, they can be used to systematically
resum logarithms of certain kinematic variables to all orders in the strong
coupling constant; for the case of $N$-jettiness, see
e.g.~Refs.~\cite{Alioli:2015toa,Alioli:2020fzf,Jouttenus:2013hs,Alioli:2021ggd,Alioli:2023rxx}.

 For hadron collisions, the most popular slicing parameters are the transverse
momentum ($q_\perp$) of the colour-less or massive coloured final-state
particles
\cite{Catani:2007vq,Bonciani:2015sha,Grazzini:2017mhc,Catani:2019iny,Kallweit:2020gcp,Catani:2020kkl,Catani:2022mfv},
and $N$-jettiness \cite{Boughezal:2015dva, Gaunt:2015pea, Boughezal:2015aha,
Boughezal:2016wmq}.\footnote{More recently, both methods have even been applied
in a combined way~\cite{Campbell:2023lcy,Neumann:2022lft}.} For both of these
slicing parameters, the factorization of cross sections in terms of soft, beam
and jet functions is well-understood
\cite{Catani:2007vq,Catani:2009sm,Stewart:2009yx,Stewart:2010tn,Becher:2010tm}.
Perturbative computations of these functions in case of $q_\perp$-slicing
parameter have been performed through N3LO in perturbative QCD
\cite{Catani:2011kr,Catani:2012qa,Gehrmann:2012ze,Gehrmann:2014yya,Li:2016ctv,Echevarria:2016scs,Luo:2019hmp,Luo:2019bmw,Gutierrez-Reyes:2019rug,Luo:2019szz,Ebert:2020yqt},
whereas for the jettiness variable only beam functions and jet functions are
known at this order
\cite{Gaunt:2014cfa,Gaunt:2014xga,Boughezal:2017tdd,Bruser:2018rad,Ebert:2020lxs,Ebert:2020unb,Baranowski:2022vcn}.
Many soft functions for $N$-jettines observables have been computed at NNLO QCD
\cite{Kelley:2011ng,Monni:2011gb,Boughezal:2015eha,Li:2016tvb,Campbell:2017hsw,Baranowski:2020xlp,Alioli:2021ggd,Bell:2023yso}.
However, at N3LO, while various contributions to the zero-jettiness soft
function have been discussed in
Refs.~\cite{Chen:2020dpk,Baranowski:2021gxe,Baranowski:2022khd,Vita:2024ypr}, the complete
result is still not available.

  In this paper, we analytically compute the one-loop correction to the
double-real emission contribution to the soft function for the zero-jettiness
variable, accounting for both $gg$ and $q \bar q$ soft final-state partons. The
result for $gg$ contribution was reported earlier in Ref.~\cite{Chen:2020dpk}.
In general, however, given the complexity of \emph{any} three-loop computation,
an independent calculation of the one-loop correction to double-real
contribution to zero-jettiness soft function is clearly warranted.

The rest of the paper is organized as follows. In Section~\ref{sect:2} we define
the soft function and introduce the notations that we use throughout the paper.
In Section~\ref{sec:calc-details} we explain how the expression for soft
currents is constructed and the integration-by-parts reduction to master
integrals is set up. In Sections~\ref{sect4} and \ref{sec:auxIntsDE} we discuss
the various methods we employed to compute the master integrals. In particular,
as we explain in Section~\ref{sec:auxIntsDE}, we found it useful to set up and
solve systems of differential equations to calculate the most challenging master
integrals. We discuss the computation of the boundary conditions for integrals
that appear in the differential equations in Section~\ref{sect6}. We present the
result for the one-loop corrections to double-real emission contributions to the
zero-jettiness soft function in Section~\ref{seq7}. We conclude in
Section~\ref{seq8}. One-loop integrals that we used for the computations
reported in this paper are collected in Appendix~\ref{app:1}. The list of all
master integrals is given in Appendix~\ref{sec:mi-list}. Results for all master
integrals as well as the expression for the one-loop correction to the
double-real emission contribution to the zero-jettiness soft function in terms
of master integrals can be found in an ancillary file provided with this paper.

  \section{The zero-jettiness soft function}
  \label{sect:2}

We begin by defining the zero-jettiness soft function. The zero-jettiness
variable in QCD is a limiting case of the so-called $N$-jettiness variable
\cite{Stewart:2009yx,Stewart:2010tn} which describes hadron collider processes
without hard QCD partons (jets) in the final state, and lepton collider
processes in which the number of hard final-state jets is exactly two. It is
defined as follows
 \begin{equation}
   \label{eq:Tsum}
   {\cal T}_0 = \sum \limits_{i=1}^{n} {\rm min} \left [ \psi_i \right],
 \end{equation}
 where the list $\psi_i$ reads
 \begin{equation}
   \psi_i  = \left \{ \frac{2 p_1\cdot k_i}{P},\frac{2p_2 \cdot k_i}{P} \right \},
 \end{equation}
and $P$ is an arbitrary normalization factor.
 
 In Eq.~(\ref{eq:Tsum}) $n$ is the total number of \emph{unresolved} final-state
partons that need to be considered in a particular order of QCD perturbation
theory, and $p_{1,2}$ are the momenta of two incoming partons that annihilate
and produce a color-less final state $X$, or the momenta of two hard partons
that are produced in a collision of an electron and a positron.
\\

The zero-jettiness soft function is constructed from integrals of soft limits of
the (various) scattering amplitudes squared, at different orders in QCD
perturbation theory over phase spaces of soft partons subject to a jettiness
constraint
$
{\cal T}_0 = \tau.
$
To be specific,  we write the perturbative expansion of the 
\emph{bare} soft function as
\begin{equation}
  S(\tau) = \delta(\tau) + S^{(1)}(\tau) +  S^{(2)}(\tau) + S^{(3)}(\tau)...,
\end{equation}
where $S^{(i)}$ is the contribution to the soft function in $i$-th order in the
perturbative expansion in QCD.

Computation of $S^{(i)}$, $i =1,2,3$, follows same rules as the calculation of
the perturbative expansion of any cross section except that {\it i}) all the
matrix elements, both real and virtual, are computed using eikonal Feynman rules
for fixed number of hard partons and {\it ii}) phase-space integration measure
contains the zero-jettiness constraint $\delta( \tau- {\cal T}_0) $ but no
energy-momentum conserving $\delta$-function.

We also note that it is convenient to deal with the case of two hard partons in
the final state, which covers the two-jet production in electron-positron
collisions. In principle, computing the soft function for hadron collisions from
the result for $e^+e^-$ could have required a non-trivial analytic continuation.
Fortunately, the zero-jettiness case is simple enough so that this does not
happen~\cite{Zhu:2020ftr,Catani:2021kcy,Chen:2020dpk} and the two results are
actually the same.

We will now discuss the phase-space integration and fully define the
contribution to the zero-jettiness soft function that we calculate in this
paper. We are interested in computing the one-loop correction to the double-real
contribution to the soft function. The double-real contribution appears for the
first time in the calculation of $S_2$; hence, the one-loop correction to it
affects $S_3$. Thus, we consider the phase space for two soft particles and use
$n=2$ in Eq.~(\ref{eq:Tsum}). Since the zero-jettiness variable requires us to
determine minima of various scalar products, it is convenient to split the phase
space into sectors where these minima are uniquely determined, and compute
contributions of these sectors separately. To this end, we introduce the Sudakov
decomposition of the momenta of two final-state soft partons. We work in the
center of mass frame of two hard partons $p_1$ and $p_2$ and write
\begin{equation}
  p_1^\mu = E \; n^\mu,  \;\;\;\ p_2^\mu = E\; \bar n^\mu.
\end{equation}
We choose $n \cdot \bar n = 2$, so that  $s = 2 p_1\cdot p_2 = 4 E^2$. We then write 
\be
k_i^\mu = \frac{\alpha_i}{2} n^\mu + \frac{\beta_i}{2} \bar n^\mu + k_{i,\perp}^\mu,
\ee
and find
\be
2 p_1\cdot k_i = 2 E \beta_i,\;\;\; 2 p_2 \cdot k_i = 2E \alpha_i.
\ee
If  we choose $P  =2 E$ in the definition of the zero-jettiness variable, we  obtain 
\be
{\cal T}_0 = \sum \limits_{i=1}^{2} {\rm min} \left [ \{ \alpha_i, \beta_i \} \right ].
\ee
There are two minimum conditions we need to resolve; hence, to identify the zero-jettiness uniquely, we use partition of unity
\begin{equation}
  \label{eq:oneTheta}
  1 = \left [ \theta( \alpha_1 - \beta_1) + \theta(\beta_1 - \alpha_1) \right ]
  \left[ \theta( \alpha_2 - \beta_2) + \theta(\beta_2 - \alpha_2) \right ].
\end{equation}
Combining this with the delta function that ensures a particular value of
zero-jettiness, we write
\begin{align}
  & \delta\left( \tau- \sum \limits_{i=1}^{2} {\rm min} \left [ \{ \alpha_i, \beta_i \} \right ] \right)
    = \nonumber\\
  & \delta\left( \tau - \beta_1 - \beta_2\right)\theta\left(\alpha_1 - \beta_1\right)\theta\left(\alpha_2 - \beta_2\right)
     + \delta\left( \tau - \alpha_1 - \alpha_2\right)\theta\left(\beta_1 - \alpha_1\right)\theta\left(\beta_2-\alpha_2\right)\label{eq:nnDelta}\\
   & + \delta\left( \tau - \beta_1 - \alpha_2\right)\theta\left(\alpha_1 - \beta_1\right)\theta\left(\beta_2-\alpha_2\right) + \delta\left( \tau - \alpha_1 - \beta_2\right)\theta\left(\beta_1 - \alpha_1\right)\theta\left(\alpha_2 - \beta_2\right).\label{eq:nnbDelta}
\end{align}

For a particular final state $f$, the one-loop correction to the double-real
contribution to the soft function is obtained by integrating the product of the
tree-level (${\cal J}^{(f)}_{0}$) and the one-loop (${\cal J}^{(f)}_{1}$)
eikonal currents over the corresponding phase space. Hence, we define
\begin{equation}
  \label{eq:softAmpSquared}
  \mathcal{M}^{(f)}\left( k_1,k_2, n,\bar{n} \right) = 2 {\rm Re} \left ( {\cal J}_{1}^{(f)} (n,\bar n, k_1,k_2) \; {\cal J}^{(f),*}_{0}(n, \bar n, k_1,k_2) \right ),
\end{equation}
where the sum over polarizations of the final-state partons is assumed.
 
The zero-jettiness constraint can be split into the same-hemisphere and the
different-hemispheres contributions, given in Eqs.~\eqref{eq:nnDelta} and
\eqref{eq:nnbDelta}, respectively. Although it appears that four different terms
need to be integrated for each of the final states $f$, we can simplify the
calculation by making use of the symmetries between final-state partons, as well
as the symmetries between two hemispheres, where appropriate. We will also use
the fact that soft eikonal currents are uniform in the soft momenta; this allows
us to factor out the dependence on the jettiness parameter $\tau$. Hence, we
write the one-loop correction to the double real-emission contribution to the
zero-jettiness soft function as
\begin{equation}
  \label{eq:SconfSplit}
  S^{(3)}_{RRV} = \tau^{-1-6\ep} \cdot \left (S^{nn}_{RRV} + S^{n \bar n}_{RRV} \right ),
\end{equation}
where 
\begin{subequations}
  \label{eq:SnnSnnbDef}
\begin{align}
  & S^{nn}_{RRV} = \sum_{\{f\}} 
    \nu_f
    \sum \limits_{{\rm spins}} 
    \int  {\rm d} \Phi^{nn}_{\theta \theta} \; \left[
    \mathcal{M}^{(f)}\left( k_1,k_2, n,\bar{n} \right) +
    \mathcal{M}^{(f)}\left( k_1,k_2, \bar{n}, n \right)
    \right],
    \label{eq:SnnDef}\\
  & S^{n\bar{n}}_{RRV} = \sum_{\{f\}} 
    \nu_f
    \sum \limits_{{\rm spins}} 
    \int  {\rm d} \Phi^{n\bar{n}}_{\theta \theta} \; \left[
    \mathcal{M}^{(f)}\left( k_1,k_2, n,\bar{n} \right) +
    \mathcal{M}^{(f)}\left( k_2,k_1, n,\bar{n}\right)
    \right],
    \label{eq:SnnbDef}
\end{align}
\end{subequations}
and $\nu_f$ are factors that are particular to the final state $f$as we describe
below. The two phase spaces that appear in the above equation read
\begin{subequations}
  \label{eq:dk1k2Phi}
  \begin{align}
    & {\rm d} \Phi^{nn}_{\theta \theta} =
      [{\rm d} k_1] \; [{\rm d} k_2]  \; \delta ( 1  - \beta_1 - \beta_2) \; \theta(\alpha_1 - \beta_1) \; \theta(\alpha_2 - \beta_2),
    \\
    &  {\rm d} \Phi^{n \bar n}_{\theta \theta} =
      [{\rm d} k_1] \; [{\rm d} k_2]  \; \delta ( 1  - \beta_1 - \alpha_2) \; \theta(\alpha_1 - \beta_1) \; \theta(\beta_2 - \alpha_2),    
  \end{align}
\end{subequations}
where $[{\rm d} k_{i}] = {\rm d}^{(d-1)}k_{i}/(2 k_{i}^{(0)}(2\pi)^{d-1})$. Note
that, as the consequence of using just the two phase spaces, we have to compute
the interferences of the one-loop and tree eikonal currents in
Eq.~\eqref{eq:softAmpSquared} for different assignments of momenta $k_{1,2}, n$
and $\bar n$.

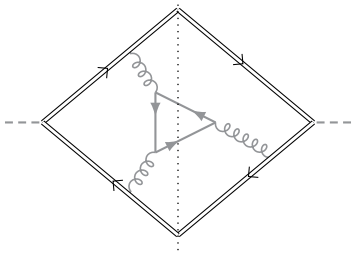
\begin{figure}[t]
  \centering
      \begin{tikzpicture}[use Hobby shortcut, scale=1]
        \coordinate (L) at (-1.8,0);
        \coordinate (R) at (1.8,0);
        \coordinate (U) at (0,1.5);
        \coordinate (D) at (0,-1.5);
        \coordinate (LU) at (-0.3,0.4);
        \coordinate (LD) at (-0.3,-0.4);
        \coordinate (RC) at (0.5,0);        
        \coordinate (k1) at (0,0.4);
        \coordinate (k2) at (0,-0.4);
        \draw[squark] (RC) -- (LU);
        \draw[squark] (LU) -- (LD);
        \draw[squark] (LD) -- (RC);
        \draw[sglu] (LU) -- (-0.65,0.95);
        \draw[sglu] (LD) -- (-0.65,-0.95);
        \draw[sglu] (RC) -- (1.3,-0.45);
        \draw[WLBC] (L)--(U);
        \draw[WLBC] (U)-- (R);
        \draw[WLBC] (R)-- (D);
        \draw[WLBC] (D)-- (L);
        \draw[higgs] (-2.3,0) -- (L);
        \draw[higgs] (2.3,0) -- (R);
        \draw [black, line width = 0.5pt, dotted] (0,-1.7) to (0,1.7);
        \path[use as bounding box] (-2.5,-1.2) rectangle (2.5,1.2);
      \end{tikzpicture}
  \caption{Example diagram which are anti-symmetric  under 
  the interchange of quark and anti-quark momenta.}
  \label{fig:SqqAsym}
\end{figure}

 We include gluons and $n_f$ quarks in our calculation. However, since we use
the Feynman gauge for both virtual and \emph{real} gluons, we must account for
final-state ghosts to remove contributions of unphysical gluon polarizations
from the final result. Hence, we write
\be
S_{RRV}^{xy} = S_{RRV}^{xy,gg}
+ S_{RRV}^{xy,c \bar c}
+ n_f S_{RRV}^{xy,q \bar q},
\ee
where $c$ is the ghost field. To construct contributions of individual final
states to $S_{RRV}$, we use $\nu_g = 1/2$, $\nu_{c} = -1$ and $\nu_q = 1$. We
also note that further simplifications in Eqs.~\eqref{eq:SnnSnnbDef} are
possible \emph{if} the matrix elements squared ${\cal M}^{(f)}$ are symmetric
under the permutations of its arguments. We find that this is indeed the case
for the gluon and ghost final states. However, it is not the case for the $q
\bar q$ final state, because diagrams of the type shown in
Fig.~\ref{fig:SqqAsym} are anti-symmetric under the interchange of the momenta
of the soft quark and the soft anti-quark. These diagrams lead to the QCD
analogy of the charge asymmetry in QED, as was recently pointed out in
Ref.~\cite{Catani:2021kcy}.

\begin{figure}[t]
  \centering
  \begin{tikzpicture}[use Hobby shortcut, scale=1]
          \coordinate (L) at (-1.5,0);
          \coordinate (R) at (1.5,0);
          \coordinate (U) at (0,1);
          \coordinate (D) at (0,-1);
          \coordinate (k1) at (0,0.4);
          \coordinate (k2) at (0,-0.4);
          \draw[sglu] (L) .. (k1) .. (R);
          \draw[sglu] (L) .. (k2) .. (R);
          \node[anchor=north east] at (k1) {\small$k_1$};
          \node[anchor=south west] at (k2) {\small$k_2$};
          \draw[WLBC] (-1.5,0.2)--(U);
          \draw[-{Stealth[scale=0.7]}] (-0.8,0.8)  --  node[pos=0.3, anchor = south] {\small$n$} (-0.3,1.0666)  {};
          \draw[WLBC] (U)-- (1.5,0.2);
          \draw[WLBC] (1.5,-0.2)-- (D);
          \draw[-{Stealth[scale=0.7]}] (-0.8,-0.8)  --  node[pos=0.3, anchor = north] {\small$\bar{n}$} (-0.3,-1.0666)  {};
          \draw[WLBC] (D)-- (-1.5,-0.2);
          \draw[higgs] (-2.3,0) -- (L);
          \draw[higgs] (2.3,0) -- (R);
          \draw [black, line width = 0.5pt, dotted] (0,-1.2) to (0,1.2);
          % Blobs
          \filldraw[fill=white, draw=black] (L) circle [radius=4mm];
          \node at (L) {\tiny \texttt{1-loop}};
          \filldraw[fill=white, draw=black] (R) circle [radius=4mm];
          \node at (R) {\tiny \texttt{tree}};
          \path[use as bounding box] (-2.5,-1.2) rectangle (2.5,1.2);
      \end{tikzpicture}
      \begin{tikzpicture}[use Hobby shortcut, scale=1]
        \coordinate (L) at (-1.5,0);
        \coordinate (R) at (1.5,0);
        \coordinate (U) at (0,1);
        \coordinate (D) at (0,-1);
        \coordinate (k1) at (0,0.4);
        \coordinate (k2) at (0,-0.4);
        \draw[sghost] (L) .. (k1) .. (R);
        \draw[sghost] (R) .. (k2) .. (L);
        \node[anchor=north east] at (k1) {\small$k_1$};
        \node[anchor=south west] at (k2) {\small$k_2$};
        \draw[WLBC] (-1.5,0.2)--(U);
        \draw[-{Stealth[scale=0.7]}] (-0.8,0.8)  --  node[pos=0.3, anchor = south] {\small$n$} (-0.3,1.0666)  {};
        \draw[WLBC] (U)-- (1.5,0.2);
        \draw[WLBC] (1.5,-0.2)-- (D);
        \draw[-{Stealth[scale=0.7]}] (-0.8,-0.8)  --  node[pos=0.3, anchor = north] {\small$\bar{n}$} (-0.3,-1.0666)  {};
        \draw[WLBC] (D)-- (-1.5,-0.2);
        \draw[higgs] (-2.3,0) -- (L);
        \draw[higgs] (2.3,0) -- (R);
        \draw [black, line width = 0.5pt, dotted] (0,-1.2) to (0,1.2);
        % Blobs
        \filldraw[fill=white, draw=black] (L) circle [radius=4mm];
        \node at (L) {\tiny \texttt{1-loop}};
        \filldraw[fill=white,draw=black] (R) circle [radius=4mm];
        \node at (R) {\tiny \texttt{tree}};
        \path[use as bounding box] (-2.5,-1.2) rectangle (2.5,1.2);
      \end{tikzpicture}
      \begin{tikzpicture}[use Hobby shortcut, scale=1]
        \coordinate (L) at (-1.5,0);
        \coordinate (R) at (1.5,0);
        \coordinate (U) at (0,1);
        \coordinate (D) at (0,-1);
        \coordinate (k1) at (0,0.4);
        \coordinate (k2) at (0,-0.4);
        \draw[squark] (L) .. (k1) .. (R);
        \draw[squark] (R) .. (k2) .. (L);
        \node[anchor=north east] at (k1) {\small$k_1$};
        \node[anchor=south west] at (k2) {\small$k_2$};
        \draw[WLBC] (-1.5,0.2)--(U);
        \draw[-{Stealth[scale=0.7]}] (-0.8,0.8)  --  node[pos=0.3, anchor = south] {\small$n$} (-0.3,1.0666)  {};
        \draw[WLBC] (U)-- (1.5,0.2);
        \draw[WLBC] (1.5,-0.2)-- (D);
        \draw[-{Stealth[scale=0.7]}] (-0.8,-0.8)  --  node[pos=0.3, anchor = north] {\small$\bar{n}$} (-0.3,-1.0666)  {};
        \draw[WLBC] (D)-- (-1.5,-0.2);
        \draw[higgs] (-2.3,0) -- (L);
        \draw[higgs] (2.3,0) -- (R);
        \draw [black, line width = 0.5pt, dotted] (0,-1.2) to (0,1.2);
        % Blobs
        \filldraw[fill=white, draw=black] (L) circle [radius=4mm];
        \node at (L) {\tiny \texttt{1-loop}};
        \filldraw[fill=white, draw=black] (R) circle [radius=4mm];
        \node at (R) {\tiny \texttt{tree}};
        \path[use as bounding box] (-2.5,-1.2) rectangle (2.5,1.2);
      \end{tikzpicture}
  \caption{Different contributions to soft amplitude squared}
  \label{fig:SggSccSqq}
\end{figure}
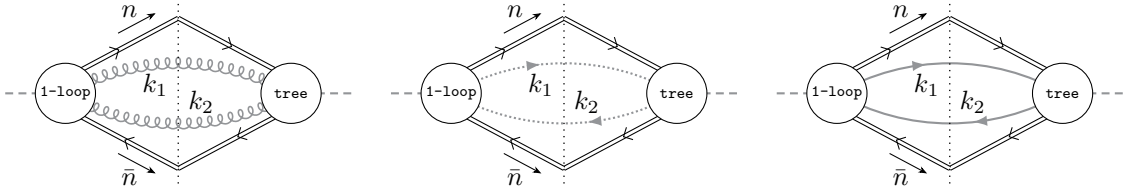

We now summarize the steps required to compute the one-loop QCD correction to
the double-real emission contribution to the zero-jettiness soft function. We
elaborate on each of these steps in the remaining sections of the paper. We
begin by constructing Born and one-loop two-parton currents using eikonal
Feynman rules,\footnote{ One-loop soft currents for $gg$ and $q \bar q$ final
states have been computed earlier, see
Refs.~\cite{Zhu:2020ftr,Czakon:2022dwk,Catani:2021kcy}. However, since we need
to integrate these currents over phase space and since this integration is
singular, we found it more convenient to construct the input for these currents
ourselves. } compute their interferences, sum over polarizations of all
final-state partons, map them onto templates in Fig.~\ref{fig:SggSccSqq} and
write the resulting contributions as integrals over phase spaces defined in
Eq.~(\ref{eq:dk1k2Phi}). To compute these integrals we rewrite them as loop-like
integrals using reverse unitarity \cite{Anastasiou:2002yz} and use
integration-by-parts technology \cite{Tkachov:1981wb,Chetyrkin:1981qh} to
express them through a small set of master integrals. However, the direct
application of integration-by-parts methods to integrals with Heaviside
functions is subtle~\cite{Baranowski:2021gxe}. As follows from
Ref.~\cite{Baranowski:2021gxe}, integration-by-parts identities for integrals
with $\theta$-functions contain integrals with some or all $\theta$-functions
replaced by $\delta$-functions. Hence, in addition to phase spaces shown in
Eq.~(\ref{eq:dk1k2Phi}), we will also require integrals over ${\rm d}
\Phi_{\delta \theta}^{nn}$, ${\rm d} \Phi_{\delta \theta}^{n \bar n}$ and ${\rm
d} \Phi_{\delta \delta}^{nn}$ where the notation should be self-explanatory. We
express all required integrals through master integrals, and then compute them
either by performing loop- and phase-space integrations directly or by setting
up and solving differential equations that these integrals satisfy. In the
following sections, we elaborate on each of these steps.

\section{Input generation and 
the reduction to master integrals}
\label{sec:calc-details}

Although it is straightforward to construct the input expression for the
interference of tree and one-loop soft currents, it requires significant amount
of bookkeeping, especially at N$3$LO QCD. Because of that, we developed a tool
chain based on the generic representation of Feynman diagrams that we have to
compute. We start with the template shown in Fig.~\ref{fig:SggSccSqq} where we
always draw a one-loop amplitude to the left of the cut and the Born amplitude
to the right of the cut. The amplitude to the right of the cut is supposed to be
complex-conjugated but it is simple to perform this operation for Born
amplitudes.

\begin{align}
  \label{eq:eikonal-feyn-rules}
  & \vcenter{\hbox{
    \begin{tikzpicture}
      \coordinate (a) at (-1,0); \coordinate (b) at (1,0);
      \draw[WLB] (a)  --  node[pos=0.15, anchor = north] {$n$} (b)  {};
      \draw[-{Stealth[scale=1]}] (-0.5,0.2)  --  node[pos=0.5, anchor = south] {$p$} (0.5,0.2)  {};
      \node[anchor=south] at (a) {\small$i$};
      \node[anchor=south] at (b) {\small$j$};
    \end{tikzpicture}
    }}
    = \frac{i \delta_{i j}}{n \cdot p + i0} \; ,
  &
  &
    \vcenter{\hbox{
    \begin{tikzpicture}
      \coordinate (a) at (-1,0); \coordinate (b) at (1,0); \coordinate (c) at (0,0.55);
      \draw[WLB] (a)  --  node[pos=0.15, anchor = north] {$n$} (b)  {};
      \draw[GLUON] (0,0.03) -- (c);
      \fill (0,0) circle (1pt);
      \node[anchor=south] at (a) {\small$i$};
      \node[anchor=south] at (b) {\small$j$};
      \node[anchor=west] at (c) {\small$\mu,a$};
    \end{tikzpicture}
    }}
    = ig n^\mu T_{ji}^a \; . &
\end{align}

To better control the input, we generate the one-loop amplitude and the tree
amplitude separately. We work with Feynman rules adapted to describe the soft
limit of QCD. To this end, in addition to conventional QCD Feynman rules, we
include \emph{eikonal} rules that parameterize the propagation of a hard parton
and its interactions with soft gluons.\footnote{For the sake of definiteness,
throughout this paper we consider the hard parton to be a quark that transforms
under the fundamental representation of $SU(3)$, although this restriction is
immaterial. Furthermore, in Section~\ref{seq7} we present the result for
arbitrary colour charges of two hard emitters provided, of course, that they are
in a colour-singlet state. }

To generate Feynman diagrams from QCD Feynman rules supplemented with additional
eikonal lines and their interactions, we developed a custom model file for the
package \texttt{DIANA}~\cite{Tentyukov:1999is} which calls
\texttt{QGRAF}~\cite{Nogueira:1991ex} and creates a suitable input for Feynman
diagrams that, together, provide soft currents, both tree and one-loop.
Furthermore, as a check, we computed one-loop and tree two-parton currents
starting from diagrams constructed from conventional QCD Feynman rules and then
treating the emitted gluons and quarks, both real and virtual, in the soft
approximation. The obtained expressions for the one-loop and tree amplitudes are
simple enough to construct their interferences and map them onto ``loop''
diagrams shown in Fig.~\ref{fig:SggSccSqq}. This latter step is useful to
simplify further calculations that we now describe. \\

The generated expression for the interference of the one-loop and tree eikonal
currents is clearly the same for $nn$ and $n \bar n$ kinematic configurations.
However, since the expression for the jettiness variable is different in the two
cases, delta-functions that ensure that the phase-space integration is performed
at fixed $\tau$ are also different. In the context of reverse unitarity, these
delta-functions are mapped onto propagator-like objects which then induce
different relations between propagators in each of the configurations. Since
these linear dependencies have to be resolved to define unique ``loop-integral''
families, these families are necessarily different for $nn$ and $n \bar n$
configurations.

 To deal with a large number of integrals, where loop and phase-space
integrations are intertwined in a complex way, it is customary to apply
integration-by-parts (IBP) reduction in the momentum space. In the context of
the computation of the zero-jettiness soft function, the application of the IBP
method requires non-standard modifications, as was first pointed out in
Ref.~\cite{Baranowski:2021gxe}. This happens because the IBP reduction of
integrals with theta-functions forces us to consider \emph{inhomogeneous} linear
relations between integrals. The inhomogeneous nature of the system of equations
that needs to be solved makes it impossible to use any of the well-tested public
codes that are used in large-scale conventional multi-loop computations for the
integral reduction. Because of that, we discuss several reduction-related points
below that are particular to the problem at hand and which were essential for
the successful application of the IBP technology to the computation of one-loop
virtual corrections to the double-real emission contributions to the
zero-jettines soft function.

Our starting point is the interference of the one-loop and tree eikonal currents
expressed through a variety of integrals with two theta-functions. We can write
integration-by-parts equations for these integrals, but these equations will
contain inhomogeneous terms which are generated when a derivative with respect
to the momentum of one of the soft partons acts on a $\theta$-function. Since
our goal is to use this system of equations to express a large number of the
original integrals in terms of a small(er) number of simpler integrals, we need
to introduce criteria to order integrals by their ``simplicity''. The most
important criterion that we use for this purpose is the \emph{total number of
theta-functions} in an integral. We will refer to this parameter as
\emph{level}. We then notice that, since the IBP equations are obtained by
computing exactly one derivative of the integrand, any IBP equation that appears
in the calculation, can be cast into the following form
 \begin{equation}
   \label{eq:ihIBPlevels}
   \sum\limits_a c_aI^L_a[\vec{n}] = \sum\limits_b c_b I^{L-1}_b[\vec{n}],
 \end{equation}
where integrals on the left-hand side have level $L$ and integrals on the
right-hand side have level $L-1$. This observation has to be contrasted with the
result of the full reduction
\begin{equation}
  \label{eq:ihIBP2mis}
  I^L_a[\vec{n}] = \sum\limits_{i=1}^{n_L} c_{ai}^L M^{L}_i[\vec{n}]
  + \sum\limits_{i=1}^{n_{L-1}} c_{ai}^{L-1} M^{L-1}_i[\vec{n}]
  +\dots
  + \sum\limits_{i=1}^{n_{0}} c_{ai}^{0} M^{0}_i[\vec{n}],
\end{equation}
which implies that an $L$-level integral will be expressed through master
integrals with levels from $L$ to zero. In writing Eq.~(\ref{eq:ihIBP2mis}) we
have assumed that there are $n_L$ master integrals $M_i^L$ at level $L$. To
order integrals further, we need to elaborate on the relative complexity of
integrals with the \emph{same} level value.

 As we have discussed, in the course of deriving the IBP relations, an integral
can move from an original level to a lower level. When this happens, an
additional partial fractioning is required since delta functions, that appear in
the integrands once $\theta$-functions are differentiated, are treated as new
propagators in the context of reverse unitarity. These new propagators often
depend linearly on the propagators that appear in original integrals and such
linear dependencies have to be resolved before one can proceed with the
reduction.
 
 The need to perform partial fractioning for each new IBP equation, except for
those that are generated at the lowest level, is one of the major complications
to efficiently setting up and solving the system of equations in this case, as
compared to more conventional cases. Our strategy was to construct templates to
generate IBP equations, perform partial fractioning, and systematically map
integrals that only contain linearly-independent propagators on integral
families defined at each level. In spite of the fact that the application of
templates to large intermediate expressions with the goal to identify and
classify integrals was crucial for the success of the reduction procedure, it
still proved to be time- and resources-consuming process.

 Similar to conventional cases, there are symmetry relations between integrals,
and the actual reduction can be performed for smaller sets of truly independent
integral families. Symmetries within a given family are based on the invariance
of integrals under linear transformations of loop and/or final-state parton
momenta but these transformations must be consistent with constraints imposed by
an observable. In the context of loop computations, such linear transformations
are applied to an integral described by a vector of powers of the propagators
$I[n_1,n_2,\dots]$ where negative values are used to describe scalar products in
numerators. A momentum transformation then leads to a re-shuffling of positive
indices and to momenta shifts in the numerators. The possible shift-identities
are universal for all integrals characterized by the same set propagators in
positive and negative powers. Following the standard terminology, these
``sub-families'' are referred to as ``sectors''. Mapping relations can be
constructed between sectors; these relations have to be ordered to uniquely
specify which sectors are kept and which ones have to mapped.

 Mapping rules between integrals can be constructed by using the method first
introduced in Ref.~\cite{Pak:2011xt}. This method relies on identifying
relations between different loop integrals by bringing their Feynman parameter
representations into a unique form so that their similarities become manifest.
To apply it to integrals with delta and theta functions, we first map all the
relevant integrals on to auxiliary loop integrals that contain all propagators
of the original integrals and \emph{additional propagators} constructed from
arguments of each delta and theta function. To ensure that propagators that
originate from delta and theta functions are not confused with conventional
propagators, we add auxiliary masses to each type of new propagators
\begin{equation}
  \theta(A) \to \frac{1}{A-m_\theta^2}, \quad
  \delta(A) \to \frac{1}{A-m_\delta^2}.
\end{equation}
We then search for symmetry relations between different integrals following the
procedure described in Ref.~\cite{Pak:2011xt}.

Following the above remarks, we introduce the global ordering of all sectors
that appear in our problem using the ${L,T_L,S_{T_L,L}}$ tuples, where $L$ is a
level, $T_L$ is a unique family label defined for the level L and $S_{T_L,L}$ is
a sector identifier for a given family. Defining the mapping rules at each
level, we construct the final list of tuples which contains unique sectors for
which IBP relations must be constructed and solved. Ordering of tuples in
complexity is fixed once and for all, and it tells us the ``correct'' order of
IBP reductions in each of the unique sectors.

 To generate IBP equations we use lists of seed integrals. It is difficult to
accurately predict the maximal powers of the numerators and denominators of
these seed integrals that are required for a successful reduction. We estimate
these powers from integrals that appear in the expression for the soft function
and we extend it to include all possible integrals that appear at the
intermediate stages. The IBP reduction starts at the maximal level ($L$=2 in our
case), where IBP identities are generated starting from the integrals in the
seed list. All integrals that appear when the IBP identities are derived are
mapped onto unique sectors. The same procedure is repeated for all simpler
sectors and lower levels, until only $L=0$ integrals are added when new IBPs are
derived. Level-zero integrals do not contain theta functions; hence, they can be
dealt with using the standard reduction methods implemented in
\texttt{Kira}~\cite{Maierhofer:2017gsa,Klappert:2020nbg}. To this end, we
generate large look-up tables which contain all possible delta-only integrals
that appear as inhomogeneous terms during the reduction.

To solve the constructed system of IBP relations, we start with the simplest
sector with $L=1$, where for the first time inhomogeneous $L=0$ contributions
appear. To deal with these inhomogeneous equations in a selected sector, we use
reduction tables for integrals from all simpler sectors and construct a linear
system readable by \texttt{Kira} where all integrals from simpler sectors are
replaced with the corresponding master integrals.

We note that \texttt{Kira}'s ability to deal with user-defined systems
tremendously simplifies and speeds up the solution of the system of linear
equations. Indeed, once the reduction tables that are needed to replace the
integrals from simpler sectors through corresponding master integrals are
provided, \texttt{Kira} can reduce the system to the minimal set of master
integrals in an efficient way using modern approaches to the IBP reduction
problem, such as finite field sampling~\cite{Klappert:2019emp}. Since this
strategy assumes that reduction results in a given sector are re-used in more
complex sectors, we prepare reduction tables for all possible integrals in a
particular sector which are limited by some powers of propagators in the
denominators and some powers of irreducible scalar products in the numerators.
It is important to note that when working with sectors at the same \emph{level},
reductions of integrals in sectors with the same number of propagators can be
done in parallel since they only depend on the reductions in simpler sectors
which are assumed to be already available.

\section{The master integrals}
\label{sect4}

Once the reduction problem is solved, master integrals have to be computed. It
follows from the above discussion that their calculation requires integration of
one-loop integrals and various factors that depend on momenta of final state
partons, over different phase spaces. We find that, in total, 61 master
integrals need to be computed; they are listed in Appendix~\ref{sec:mi-list}.
All one-loop integrals, needed to compute the master integrals for the soft
function, contain two-, one- or no eikonal lines. All of them, except the
five-point integrals, can be easily expressed in terms of hypergeometric
functions; the results can be found in Appendix~\ref{app:1}. Furthermore, there
is a large number of master integrals for which integration over phase spaces of
two soft partons can also be performed with a relative ease.

To show an example of how such computations are performed, consider master
integral defined in Eq.~(\ref{eqb.56}). It reads
\begin{equation}
  \label{eq:I56}
  I_{56} = \int \frac{ {\rm d} \Phi_{\theta \theta}^{n \bar n} } {\left(k_1\cdot k_2\right)  ( k_2 \cdot n) ( k_{12} \cdot  \bar{n} )}\; \pent{2,1}{10111},  
\end{equation}
where $k_{12} = k_1 + k_2$ and 
 $\pent{2,1}{10111}$ can be found in Appendix~\ref{app:1}. We write it here one more time
for the ease of reference 
\begin{equation}
  \label{eq:P21x10111}
  \pent{2,1}{10111}   = 
  \frac{i \Omega^{(d-2)}}{4 (2\pi)^{d-1}}  e^{-i \pi \ep}  \left[ \frac{ \Gamma(1-\ep) \Gamma(1+\ep)}{\ep} \right ]^2
  ( \alpha_1 \beta_{12} )^{-1-\ep} {}_2F_1\left ( 1+\ep,-\ep,1-\ep;\frac{\beta_1}{\beta_{12}} \right ),
\end{equation}
where $\Omega^{(n)}=2\pi^{n/2}/\Gamma(n/2)$ denotes the surface of a unit sphere
embedded in $n$ dimensions.

To proceed with the calculation of the integral $I_{56}$, we begin by
integrating over the relative azimuthal angle between transverse momenta of the
two gluons. Since $ \alpha_1 \beta_2 > \alpha_2 \beta_1$ for $n \bar n$
integrals, we find
\begin{equation}
  \label{eq:k1k2OmegaInt}
  \int \frac{{\rm d} \Omega^{(d-2)}}{\Omega^{(d-2)}} \frac{1}{\left(k_1\cdot k_2\right)}
  = \frac{2}{\alpha_1 \beta_2} {}_2F_1\left (1,1+\ep,1-\ep;\frac{\alpha_2 \beta_1 }{\alpha_1 \beta_2 } \right ).
\end{equation}
The required integral becomes 
\be
\begin{split} 
I_{56}& =
2 i \left (
\frac{\Omega^{(d-2)}}{4 (2\pi)^{d-1}}  \right )^3 e^{-i \ep \pi}
\left [ \frac{\Gamma(1-\ep) \Gamma(1+\ep)}{\ep}  \right ]^2 \int \prod
        \limits_{i=1}^{2} {\rm d} \alpha_i {\rm d} \beta_i (\alpha_i \beta_i)^{-\ep} 
\\
& \times \delta(1-\beta_1 - \alpha_2)
\theta(\alpha_1 - \beta_1) \theta(\beta_2 - \alpha_2) 
( \alpha_1 \beta_{12} )^{-1-\ep} (\alpha_{12} \beta_2^2 \alpha_1 )^{-1} 
\\
& \times 
  {}_2F_1\left (1,1+\ep,1-\ep;\frac{\alpha_2 \beta_1 }{\alpha_1 \beta_2 } \right ).
 {}_2F_1\left (-\ep,1+\ep,1-\ep;\frac{\beta_1}{\beta_{12}} \right )
\end{split} 
\ee
We change variables $\alpha_{1}\to r_{1}$,
$\beta_2 \to r_2$, where 
\be
\alpha_1 = \frac{\beta_1}{r_1}, \;\;\;\; \beta_2 = \frac{\alpha_2}{r_2},
\ee
and use the $\delta$-function to eliminate $\beta_1$.  We  find
\begin{equation}
  \label{eq:I56noTheta}
  \begin{split} 
    I_{56} & =
             2 i \left (
             \frac{\Omega^{(d-2)}}{4 (2\pi)^{d-1}}  \right )^3 e^{-i \pi \ep}
             \left [ \frac{\Gamma(1-\ep) \Gamma(1+\ep)}{\ep}  \right ]^2
             \int \limits_{0}^{1} {\rm d} r_1 \; {\rm d} r_2 \; {\rm d} \alpha_2
    \\
           & \times
             \frac{\alpha_2^{-1-2 \ep} (1-\alpha_2) ^{-1-3\ep} r_1^{1+2 \ep} r_2^{1+2\ep} 
             }{ (1-\alpha_2(1-r_1) ) ( \alpha_2 + (1-\alpha_2) r_2)^{1+\ep}}
    \\
           & \times {}_2F_1\left(1,1+\ep,1-\ep;r_1 r_2\right)
             {}_2F_1 \left (-\ep,1+\ep,1-\ep;\frac{(1-\alpha_2) r_2 }{\alpha_2 + (1-\alpha_2) r_2 } \right ).
  \end{split}
\end{equation}
To isolate singularities in this integral, we note that hypergeometric functions
in Eq.~\eqref{eq:I56noTheta} do not contain isolated singularities. In fact,
${}_2F_1 (1,1+\ep,1-\ep;r_1 r_2) \sim 1/(1-r_1 r_2)$ and ${}_2F_1 \left
(-\ep,1+\ep,1-\ep;(1-\alpha_2) r_2/(\alpha_2 + (1-\alpha_2) r_2) \right )$ has a
logarithmic singularity at $\alpha_2 = 0$. To isolate non-integrable
singularities, we perform partial fractioning of the prefactor in
Eq.~\eqref{eq:I56noTheta} and find
\begin{equation}
  \label{eq:I56pf}
  \begin{split} 
    & \frac{ r_1 r_2}{ \alpha_2 (1-\alpha_2) (1-\alpha_2(1-r_1) ) ( \alpha_2 + (1-\alpha_2) r_2)}
    \\
    &   = \frac{r_1}{\alpha_2} + \frac{r_2}{1-\alpha_2} - \frac{r_2(1-r_1)^3}{(1-r_{1} r_2) (1-\alpha_2(1-r_1))}
      - \frac{r_1 (1-r_2)^3}{(1-r_{1} r_2) (r_2+\alpha_2(1-r_2)}.
  \end{split}
\end{equation}
If the last two terms in the right hand side of Eq.~\eqref{eq:I56pf} are used
instead of the expression in the left hand side of this equation in
Eq.~\eqref{eq:I56noTheta}, integrations over $\alpha_2$ and $r_{1,2}$ can be
performed upon expanding the integrand in $\ep$ since it does not contain
non-integrable singularities. We use \texttt{HyperInt} \cite{Panzer:2014caa} to
perform the required integrations.

The second term in the right hand side of Eq.~\eqref{eq:I56pf} has a singularity
at $\alpha_2=1$. However, $\alpha_2 = 1$ is a regular point of the
hypergeometric function
\be
G(\alpha_2,r_2) = {}_2F_1 \left (-\ep,1+\ep,1-\ep;
\frac{ (1-\alpha_2) r_2}{(\alpha_2 + (1-\alpha_2) r_2) }\right ),
\label{eq4.8a}
\ee
and of an $\ep$-dependent term $( \alpha_2 + (1-\alpha_2) r_2)^{-\ep}$ in
Eq.~\eqref{eq:I56noTheta}. Hence, focusing on the second term in
Eq.~\eqref{eq:I56pf} and the $\alpha_2$-integration we write
\be
\begin{split}
  & \int \limits_{0}^{1}   \; 
            \frac{
            {\rm d} \alpha_2 \;
            \alpha_2^{-2 \ep} (1-\alpha_2) ^{-1-3\ep}  
             }{  ( \alpha_2 + (1-\alpha_2) r_2)^{\ep}}
    G(\alpha_2,r_2) =
     \\
  &    \int \limits_{0}^{1}  \; 
       \frac{ 
       {\rm d} \alpha_2 \;
       \alpha_2^{-2 \ep}}{ (1-\alpha_2) ^{1+3\ep} }
        \Bigg \{  1 
 + 
        \Bigg  [
            \frac{G(\alpha_2,r_2)}{ ( \alpha_2 + (1-\alpha_2) r_2)^{\ep}}
       - 1
     \Bigg  ] 
     \Bigg \}.
\end{split}
\end{equation}
The first term in the curly brackets on the right hand side of the above
equation can be immediately integrated, and the term in the square brackets
expanded in $\ep$ as it is not singular at $\alpha_2 = 1$. The following
integrations over $r_{1,2}$ are also not singular and can be easily performed
using \texttt{HyperInt}.

We continue with the discussion of the first term in \eqref{eq:I56pf}.
Substituting it into the rest of Eq.~\eqref{eq:I56noTheta}, changing the
integration variable $\alpha_2 \to u$, where
\begin{equation}
  \label{eq:I56al2toU}
  \alpha_2 =  \frac{r_2 (1-u)}{r_2 + u (1-r_2)},
\end{equation}
and integrating over $r_1$, we obtain the following integral to compute
\begin{align}
  \label{eq:I56part1}
  J \sim \frac{\Gamma(2 + 2 \ep)}{\Gamma(3 + 2 \ep)} \int\limits_{0}^1
  & \dm u \dm r_2 \; r_2^{-\ep} (1 - u)^{-1 - 3 \ep} u^{-3 \ep} ( u + r_2 (1-u))^{-1 + 6 \ep}\nonumber\\
  & {}_2F_1\left( 1, -2 \ep, 1 - \ep; u \right)
    {}_3F_2\left( \{1, 1 + \ep, 2 + 2 \ep \}, \{1 - \ep, 3 + 2 \ep\}; r_2\right).
\end{align}
This integral is divergent at  $u\to 1$.  To extract this 
divergence, we note that the following equation holds
\be
\begin{split}
  \label{eq:I56trF21u2ub}
  & (1 - u)^{-1 - 3 \ep} {}_2F_1\left( 1, -2 \ep, 1 - \ep; u \right) =   (1 - u)^{-1 - 2 \ep} u^{\ep} \frac{2\Gamma^2(1-\ep)}{\Gamma(1-2\ep)}
  \\
  & 
    - (1 - u)^{-1 - 3 \ep} 
    - (1 - u)^{- 3 \ep}\frac{ 2 \ep^2}{\Gamma(2-\ep)}  {}_2F_1\left( 1, 1-2 \ep, 2 - \ep; 1-u \right).
\end{split}
\ee
To prove it, we first use the standard identity that relates hypergeometric
functions with arguments $u$ and $1-u$ and, then employ one of the Gauss'
relations
\begin{equation}
  \label{eq:F21contig}
  {}_2F_1\left( a, b, c; 1-u \right) = {}_2F_1\left( a-1, b, c; 1-u \right) + \frac{b}{c}(1-u) {}_2F_1\left( a, b+1, c+1; 1-u \right)
\end{equation}
to shift parameters of one of the hypergeometric functions.

When we use Eq.~(\ref{eq:I56trF21u2ub}) in Eq.~(\ref{eq:I56part1}) the first two
terms on the r.h.s. of Eq.~(\ref{eq:I56trF21u2ub}) can be integrated over $u$ in
terms of hypergeometric functions and, since the remaining integration over
$r_2$ is non-singular, it can be performed expanding the integrand in powers of
$\ep$ and integrating using \texttt{HyperInt}. The third term on the r.h.s. of
Eq.~(\ref{eq:I56trF21u2ub}) does not generate singularities when integrated over
$u$ and $r_2$ in Eq.~(\ref{eq:I56part1}). Thus, we integrate it with the help of
\texttt{HyperInt} after expanding in $\ep$.

Finally, putting everything together, we find the following result for the
integral defined in Eq.~\eqref{eq:I56}
\begin{equation}
  \label{eq:I56exp}
  \begin{split} 
    & I_{56} = 2 i \left (\frac{\Omega^{(d-2)}}{4 (2 \pi)^{d-1}} \right )^3 e^{-i \pi \ep} \Bigg [  
      -\frac{1}{\ep^3}
      +\frac{1}{\ep^2} \left ( -4 \zeta_3 -\frac{\pi ^2}{6}+5 \right )
    \\
    & + \frac{1}{\ep} \left ( -5 \zeta_3  +\frac{7 \pi ^2}{3}
      -\frac{\pi ^4}{30}-17 \right )
      +\frac{40 \pi ^2 \zeta_3 }{3}+117 \zeta_3 -200 \zeta_5 -\frac{34 \pi^2}{3}-\frac{19 \pi^4}{180}+41
    \\
    & +\ep \left( 302 \zeta_3^2 -\frac{23 \pi ^2 \zeta_3}{6} -643 \zeta_3
      +15 \zeta_5 +\frac{601 \pi ^4}{120}+43 \pi^2-\frac{5339 \pi^6}{11340}
      -29\right) 
      +{\cal O}(\ep^2) \Bigg ],
  \end{split} 
\end{equation}
where we performed an expansion in $\ep$ up to ${\cal O}(\ep)$ when weight-six
contributions appeared for the first time. Although the absolute majority of the
required integrals can be computed following similar steps, there are several
integrals for which doing this appears to be difficult. As an example, we
consider the integral in Eq.~(\ref{eqb.49}) which we also display here for
convenience
\be
I_{49} = \int {\rm d} \Phi^{n \bar n }_{\theta \theta} \;
\frac{\pent{1,2}{11011}}{\left(k_1 \cdot k_2\right)},
\label{eq3.11}
\ee
In Eq.~(\ref{eq3.11}) $\pent{1,2}{11011}$ is a one-loop four-point function that can be found in Appendix~\ref{app:1}. It reads
\be
\begin{split} 
  \pent{1,2}{11011} & = \int \frac{{\rm d}^d l}{(2 \pi)^d} \; \frac{1}{ [ l \cdot n ]  \; [ l^2 ]  [ (l - k_{12})^2 ]
                      [ (k_{12} - l ) \cdot \bar n ] } =  \frac{i \Omega^{(d-2)}}{4 (2\pi)^{d-1}}  e^{-i \pi \ep} \;
  \\
                    &   \times \frac{\Gamma^3(1-\ep) \Gamma^2(1+\ep)}{
                      \ep^2 \Gamma(-2\ep) \Gamma(2+\ep) \; (\alpha_{12} \beta_{12} )^{1+\ep} } {}_2F_1\left ( 1+\ep,1+\ep,2+\ep;1 - \frac{k_{12}^2}{\alpha_{12} \beta_{12} }
                      \right ),
\end{split}
\label{eq3.12}
\ee
The difficulty with evaluating the integral in Eq.~(\ref{eq3.11}) is related to
the appearance of the invariant mass of the two gluons $k_{12}^2$ in the
argument of the hypergeometric function in Eq.~(\ref{eq3.12}). Since $k_{12}^2$
depends on the relative angle between transverse momenta of the two gluons, the
integration over this azimuthal angle becomes very complicated. It is quite
possible that, with some effort, one can find a way to expand this integral in
powers of $\ep$ or derive a suitable Mellin-Barns representation for this
integral, but we decided not to pursue this effort. Instead, we have opted for
constructing a system of differential equations for the remaining integrals by
adding auxiliary parameters to the problem. Below we discuss how this can be
done.

\section{Differential equations for master  integrals}
\label{sec:auxIntsDE}
In this section we explain how to set up differential equations for master
integrals that appear in the calculation of the one-loop correction to the
double-real emission contribution to the zero-jettiness soft function. The key
observation is that we can replace $\theta$-functions in all master integrals
that we need to compute with integrals of $\delta$-functions that depend on the
argument of the original theta function and an auxiliary parameter. We then use
these parameters to derive and solve the differential equations for descendants
of the original master integrals that contain these modified delta functions.

To illustrate this point further, we consider a $\theta \delta$ integral
\be
I_{\theta \delta} = \int {\rm d} \Phi^{nn}_{\theta \delta} f(k_1,k_2).
\ee
Since this is an $nn$-integral, it contains a theta function $\theta(\alpha_1 -
\beta_1)$. We then use the integral representation
\cite{Baranowski:2020xlp}\footnote{See also Section 4.2.2 in
Ref.~\cite{Angeles-Martinez:2018mqh}.}
\be
\theta(\alpha_1 - \beta_1) = \alpha_1 \int \limits_{0}^{1} {\rm d} z   \; \delta ( \alpha_1 z - \beta_1),
\ee
and write 
\be
I_{\theta \delta} = \int \limits_{0}^{1} {\rm d} z  \;J_{\theta \delta}(z),
\ee
where
\be
J_{\theta \delta}(z) = \int {\rm d} \Phi^{nn}_{ \delta_z \delta } \; \alpha_1  \;  f(k_1,k_2).
\ee
The modified phase space  in the above equation is defined as
\be
   {\rm d} \Phi^{nn}_{\delta_z \delta} = [{\rm d}k_1]
   \; [{\rm d} k_2] \; \delta(1 - \beta_1 - \beta_2) \;
    \delta (\alpha_1 z - \beta_1) \; \delta( \alpha_2 - \beta_2).
\ee
Since there is a parameter $z$ in an integral with the $\delta$-functions, we
can use reverse unitarity to derive differential equations for $J_{\theta
\delta}(z)$ integrals. Exactly the same can be done for $\theta
\theta$-integrals where the only difference is that we will have to rewrite
\emph{each} $\theta$-function as an integral of a $\delta$-function so that
resulting auxiliary integrals depend on two parameters $z_{1,2}$.

Although $\delta \delta$ integrals are, in principle, simpler, not all of them
are simple enough to allow seamless direct computation. For this reason, it is
useful to set up a system of differential equations also for $\delta \delta$
integrals. Since the trick that we used for $\theta \delta$ and $\theta \theta$
integrals cannot be used for $\delta \delta$ integrals, we need to introduce an
auxiliary parameter in a different way. Specifically, we consider a $\delta
\delta$-integral
 \be
  I_{\delta \delta}  = \int {\rm d} \Phi_{\delta \delta}^{nn} \; f(k_1,k_2).
     \ee
and 
write it as 
\be
I_{\delta \delta}
= \int \limits_{0}^{1} 
{\rm d} x \; J_{\delta \delta} (x),
\label{eq5.7}
\ee
 where 
 \be
 J_{\delta \delta} (x) = \int {\rm d} \Phi_{\delta \delta}^{nn} \; f(k_1,k_2)  \; \delta( k_{12}^2 - x).
 \ee
 The parameter $x$ fixes the invariant mass of the two soft gluons
\cite{Chen:2020dpk}. We note that the integration boundaries in
Eq.~(\ref{eq5.7}) are easily established using the Sudakov parametrization for
the final-state gluons' momenta. Similarly to the case of $z$-dependent
auxiliary integrals discussed earlier, it is straightforward to derive
differential equations for $J_{\delta \delta}(x)$ using reverse unitarity.

To summarize, following the above discussion we conclude that all master
integrals that we require to calculate the one-loop corrections to the
double-real emission contribution to the zero-jettiness soft function can be
written in the following way
\begin{subequations}
  \label{eq:auxIntDef}
  \begin{align}
    \vec{I}^{\,\delta\delta} & = \int\limits_{0}^1 \dm x \; \vec{J}^{\,\delta\delta}(x)
                               = \int\limits_{0}^1 \dm x \; R^{\delta\delta}(\ep,x) \; \vec{f}^{\,\delta\delta}(x),
                               \label{eq:auxIntDefDD}\\
    \vec{I}^{\,\theta\delta} & = \int\limits_{0}^1 \dm z \; \vec{J}^{\,\theta\delta}(z)
                               = \int\limits_{0}^1 \dm z   \; R^{\theta\delta}(\ep,z)
                               \; \vec{f}^{\,\theta\delta}(z),
                               \label{eq:auxIntDefDT}\\
    \vec{I}^{\,\theta\theta} & = \int\limits_{0}^1 \dm z_1 \int\limits_{0}^1\dm z_2 \; \vec{J}^{\,\theta\theta}(z_1,z_2)
                               = \int\limits_{0}^1 \dm z_1 \int\limits_{0}^1 \dm z_2 \; R^{\theta\theta}(\ep,z_1,z_2) \; \vec{f}^{\,\theta\theta}(z_1,z_2),
                               \label{eq:auxInDfTT}
  \end{align}
\end{subequations}
where ${\vec J}^{\delta \delta, \theta \delta, \theta \theta}$ are the
parameter-dependent descendants of the original master integrals whereas ${\vec
f}^{\delta \delta, \theta \delta, \theta \theta}$ are $x$-, $z$- and
$z_{1,2}$-dependent master integrals. The functions $R^{\delta \delta, \theta
\delta, \theta \theta}$ are the reduction coefficients which arise when $\vec J$
integrals are expressed in terms of $\vec f$ integrals; they are rational
functions of $\ep$ and $x$, $z$ or $z_{1,2}$. Since ${\vec J}$ integrals do not
contain $\theta$-functions, the reduction of $\vec J$ to $\vec f$ , is performed
using the standard \texttt{Kira} interface.

As we already mentioned, $\vec f$ integrals satisfy differential equations which
can be constructed by differentiating w.r.t $x$, or $z$, or $z_{1,2}$, and then
reducing new $\vec f$ integrals back to the original set. We note that the set
of $\vec f$ integrals should be large enough so that the system of differential
equations closes.

Once the integral representations of $\vec I$ integrals shown in
Eq.~\eqref{eq:auxIntDef} and the differential equations for $\vec f$ integrals
are derived, we solve these differential equations and determine $\vec f$
integrals provided that an appropriate set of boundary conditions is known.
Since these integrals are sufficiently complex, we can only compute them
expanding in $\ep$. We emphasize that such results for master integrals are
valid in the bulk of the corresponding integration intervals but not at the
boundaries where the integrals are singular. For this reason, it is in general
\emph{not} possible to use the $\ep$-expanded results for the master integrals
to integrate over $x,z$ and $z_{1,2}$ because the corresponding integrals
diverge at the integration boundaries.

To explain how, in spite of this problem, the $\vec f$ integrals are computed
and then integrated over auxiliary parameters, we note that a basis for these
integrals can be chosen that brings the system of differential equations to the
so-called $\ep$-form~\cite{Henn:2013pwa}. We will refer to this new basis of
$\vec f$ integrals as $\vec g$ integrals.

For integrals that depend on a single variables, Eq.~\eqref{eq:auxIntDefDD} and
Eq.~\eqref{eq:auxIntDefDT}, the differential equations in the $\ep$-form read
\begin{equation}
  \label{eq:epFormDEvar1}
  \frac{\partial \vec{g}}{ \partial y}  = \ep A(y) \vec{g}, \quad \vec{f} = T \vec{g}.
\end{equation}

For the case of two variables Eq.~\eqref{eq:auxInDfTT} we obtain a system of
coupled partial differential equations
\begin{equation}
  \label{eq:epFormDEvar2}
  \frac{\partial \vec{g}(y_1,y_2)}{ \partial y_1}  = \ep A_1(y_1,y_2) \vec{g}(y_1,y_2),
  \quad
  \frac{\partial \vec{g}(y_1,z_2)}{ \partial y_2}  = \ep A_2(y_1,y_2) \vec{g}(y_1,y_2), \quad \vec{f} = T \vec{g}.
\end{equation}
We note that we changed variables $x=y^2$ and $z=y^2$ for single-variable
integrals and $z_i=y_i^2$ for integrals in \eqref{eq:auxInDfTT} to write the
differential equations Eqs.~(\ref{eq:epFormDEvar1},\ref{eq:epFormDEvar2}). For
all the required cases the transformation matrices $T$ were found using
\texttt{CANONICA}~\cite{Meyer:2017joq}.

It turns out that the single-variable differential equations
Eq.~\eqref{eq:epFormDEvar1} in our case are such that the general solution
expanded in $\ep$ can be immediately written in terms of harmonic
polylogarithms~\cite{Remiddi:1999ew}. For the two-variable case
Eq.~\eqref{eq:epFormDEvar2}, the general solution of the system is expressed
through the generalized polylogarithms~\cite{Goncharov:1998kja}.\footnote{We
note that the alphabet for this case reads $\mathcal{A} = \left\{ y_1,1\pm y_1 ,
y_2 , 1 \pm y_2, y_1 \pm y_2, 1 \pm y_1 y_2 \right\}$.} To construct particular
solutions we need to determine boundary conditions. We do this by calculating
selected integrals around chosen singular points of the differential equations.
To this end, consider a singular point $y_0$. Since this is a regular-singular
point of the relevant systems of differential equations, the expansion of the
integrals around it can be written as follows
\be
\vec g = U_{y\to y_0} \vec C_{y_0},
\label{eq5.12}
\ee
where $\vec C_{y_0}$ is the vector of constants and the matrix $U_{y \to y_0}$
reads\footnote{Expansion of the matrix $U_{y \to y_0}$ around a singular point
may include logarithms of $y - y_0$. We have found, empirically, that these
logarithms are absent in the considered case.}
\be
  \label{eq:UexpAnz}
  U_{y\to y_0} = \sum\limits_{\lambda} (y - y_0)^{\lambda \ep}\sum\limits_{n=0}^{\infty}c_{\lambda,n} (y-y_0)^n.
\ee
All possible values of parameters $\lambda$ as well as recursive relations
between \emph{matrix} coefficients $c_{\lambda,n}$ can be obtained from the
differential equations Eq.~\eqref{eq:epFormDEvar1} which the matrix $U_{ y \to
y_0}$ satisfies. We note that the rational matrix $A$ in that equation should
also be expanded in series around $y = y_0$. The resulting solutions contain a
few constants that need to be determined separately. To this end, one computes
the expansion of the original $\vec f$ integrals at $y = y_0$ by calculating
$\vec f = T \vec g$. The constants $\vec C_{y_0}$ can then be determined from
the explicit computation of a few ``branches'' of $\vec f$ integrals in the
vicinity of $y = y_0$. This step completes the determination of particular
solutions for single-variable integrals $\vec f$.

Once integrals $\vec f$ are known as an expansion in $\ep$, their linear
combinations with rational coefficients need to be integrated over $x$ or $z$
from zero to one, to recover the original master integrals, c.f.
Eq.~(\ref{eq:auxIntDef}). As it turns out, these integrations, are singular at
the boundaries. The singularities can appear both in matrices $R$ and in the
integrals $\vec f$ themselves. To deal with these singularities, we need to
construct $\ep$-exact approximations to integrands in
Eqs.~(\ref{eq:auxIntDefDD},\ref{eq:auxIntDefDT}) around singular points such
that differences between actual and approximate integrands are integrable in the
whole interval $[0,1]$ for $\ep = 0$. Below we explain how this can be done
using the $\ep$-expanded solutions in the bulk of the interval provided that
they are known to a sufficiently high $\ep$-order.

The challenge is to construct an $\ep$-exact solution of the differential
equation in the vicinity of the singular point $y = y_1$ from the available
$\ep$-expanded solution valid at $y \ne y_1$. This is an ``inverse'' problem
to what we did when we constructed particular solutions from the approximate
computation of integrals at $y = y_0$. Hence, if we need to calculate the
$\ep$-resummed approximation to the solution of the differential equations at a
point $y = y_1$, we write an Ansatz similar to Eqs.~(\ref{eq5.12},
\ref{eq:UexpAnz}) with $y_0$ replaced by $y_1$, expand the approximate result in
$\ep$ and the exact one in $y - y_1$, and determine the vector of constants
$\vec C_{y_1}$. Once this is accomplished, we can compute the expansion of
$\ep$-exact integrals in $y-y_1$ to any order using recurrence relations derived
from the differential equations, thereby determining the required subtraction
terms.

We now explain how to do this in an efficient way. We start by writing the
vector of unknown constants $\vec C_{y_1}$ as
\begin{equation}
  \label{eq:fixCgU}
  \vec{C}_{y_1} = U_{y\to y_1}^{-1} \vec{g},
\end{equation}
where $\vec{g}$ is known as an expansion in $\ep$. In principle, we can compute
the matrix $U_{y \to y_1}$ following the steps discussed in connection with the
evaluation of the matrix $U_{y \to y_0}$. However, it turns out to be more
efficient to derive and solve the differential equation satisfied by $U^{-1}_{y
\to y_1}$ instead of solving an equation for $U_{y \to y_1}$ and then inverting
the matrix. We note that the equation for the inverse matrix is easily derived
from an equation that the matrix $U$ satisfies.

We use these results to enable the calculation of the original master integrals
$\vec I$. Accounting for the change of variables $x,z \to y^2$, we write
\begin{align}
  \label{eq:oneDimIntSubtr}
  \vec{I} & = \int\limits_{0}^{1} \dm y \left[(2 y) \; R \; T \vec{g} - \vec{S} \right]
            + \int\limits_{0}^{1} \dm y \; \vec{S}. 
\end{align}
The subtraction term $\vec S$ removes all singular terms of the expression $2 y
RT \vec g$ in both $y\to 0$ and $y\to 1$ limits; it is obtained from expansions
of the corresponding $U$-matrices around singular points using representation
shown in Eq.~\eqref{eq:UexpAnz}. It reads
\begin{equation}
  \label{eq:subtr1dDef}
  \vec{S} = (2 y) R T \left[ U_{y\to 0}\vec{C}_{y\to 0} + U_{y\to 1}\vec{C}_{y\to 1} \right].
\end{equation}
By definition the integration of the first term of Eq.~(\ref{eq:oneDimIntSubtr})
converges on the interval $y \in [0,1]$ and we perform it using the package
\texttt{HyperInt}~\cite{Panzer:2014caa}.

The integration of the subtraction term is elementary since it only contain
powers of $y$ and $(1-y)$. The only difficulty here is a need for a deep
expansion of vectors $\vec{C}_{y\to 0}$ and $\vec{C}_{y\to 1}$ in $\ep$.
However, using procedures explained above it is possible to construct such
expansions provided that solutions to differential equations in an
$\ep$-expanded form are known to sufficiently high orders in $\ep$.

We treat integrals that depend on two integration variables in a similar way,
except that the boundary conditions now are series in two variables. This leads
to certain changes in the described algorithm. We start from the system of
differential equations in a canonical form,
\begin{equation}
  \label{eq:deSysz1z2}
  \frac{\partial}{\partial y_1} \vec{g}(y_1,y_2) = M_1 \vec{g}(y_1,y_2),
  \quad
  \frac{\partial}{\partial y_2} \vec{g}(y_1,y_2) = M_2 \vec{g}(y_1,y_2),
\end{equation}
where 
\be
M_i = \ep A_i(y_1,y_2),
\ee
and solve the first differential equation. We then find 
\be
\vec g =  U 
\vec b, 
\label{eq5.19}
\ee
where $\vec b$ is now a \emph{function} of the second variable. Substituting
Eq.~(\ref{eq5.19}) into the differential equation for $y_2$, c.f.
Eq.~(\ref{eq:deSysz1z2}), we obtain
\begin{equation}
  \label{eq:bDEy2}
  \frac{\partial}{\partial y_2} \vec{b} = U^{-1}\left[ M_2 U - \partial_{y_2} U\right] \vec{b} = M\; \vec{b}.
\end{equation}
By construction, this equation does not depend on $y_1$ and, therefore, can be
solved following what has been discussed earlier. Hence, we bring this equation
to a canonical form with the help of an additional transformation $\vec{b} =
T^\prime \vec{b}^{\prime}$ and write its solution as
\be
\vec{b}^{\prime} 
= V \vec{d}, 
\ee
where $\vec d$ is a vector of constants. Finally, the solution of the original
two-variable system is written as follows
\begin{equation}
  \label{eq:fSolEpExp}
  \vec{f}(y_1,y_2)
  = T \; \vec{g}(y_1,y_2)
  = T \; U(y_1,y_2) \; \vec{b}(y_2)
  = T \; U(y_1,y_2) \;  T^{\prime}  \; V(y_2) \vec{d}.
\end{equation}
The constants $\vec{d}$ can then be determined by comparing the expansion of the
integrals $\vec f$ around chosen singular points with the expansion of matrices
on the right hand side of the above equation. We note that, in variance with two
single-variable cases, matrix $U(y_1,y_2)$ contains generalized Goncharov
polylogarithms, whereas matrix $V$ depends on harmonic polylogarithms only.

Finally, to construct the subtraction terms which are required to enable
integration over $y_{1,2}$, we proceed in the same way as in the single-variable
case. The difference, however, is that we need to construct generalized
expansions in two variables. Since we need $\vec f(y_1,y_2)$ integrals to
compute $\theta \theta$ integrals and since these integrals only appear in an $n
\bar n$ configuration which is less singular in comparison to $nn$, it appears
that a fairly small number of such subtraction terms needs to be constructed. To
manipulate and integrate generalized polylogarithms that depend on two
variables, we use their implementation in \texttt{HyperInt}.

\section{Boundary conditions}
\label{sect6}

To compute integrals by solving the differential equations, we require boundary
conditions. The point for which the boundary conditions are calculated is chosen
in such a way that computation of integrals simplifies. To understand for which
values of $x$, $z$ and $z_{1,2}$ such simplifications happen, we need to
consider the dependence of phase spaces and various scalar products on these
parameters. The most complicated scalar product is $2 k_1 k_2$ because of its
dependence on the azimuthal angle between momenta of the two gluons $k_{1,2}$.
Hence, an appropriate choice of the kinematic point should simplify this
dependence.

We begin with the discussion of $\delta \delta$ integrals.  In this case, 
the phase-space reads 
\be
   {\rm d} \Phi =  [{\rm d} k_1 ] \; [{\rm d} k_2 ]
       \delta( 1 - \beta_1 - \beta_2 ) \delta(\alpha_1 - \beta_1)
       \delta(\alpha_2 - \beta_2) \delta ( 2 k_1 k_2 - x).
       \label{eq6.1}
\ee
Since for $\delta \delta$ integrals 
\be
x = 2 k_1 k_2 = 2 \beta_1 \beta_2 ( 1- \cos \varphi_{12} ), 
\ee
we can trade the integration over $\cos \varphi_{12}$ for the integration over $x$. We find 
\be
{\rm d} \Phi = 
\left ( \frac{ \Omega^{(d-2)}}{4 (2 \pi)^{d-1}} \right )^2
N_x  \; x^{-1/2-\ep} \left ( 4 \beta_1 ( 1- \beta_1) - x
\right )^{-1/2-\ep} \; {\rm d} x \; {\rm d} \beta_1,
\ee
where 
\be
N_x = \frac{2^{4 \ep} 
\Gamma(1 - 2 \ep)}{\Gamma(1/2 - \ep)^2},
\ee
is the normalization factor. 

A convenient point to compute the boundary conditions is $x \to 1$. Indeed,
since the argument of the square root in the above equation needs to be
positive, the integration region over $\beta_1$ is given by the following
equation
\be
 \frac{1}{2} ( 1 - \sqrt{1-x} ) < \beta_1 < \frac{1}{2} ( 1 + \sqrt{1 - x} ).
 \label{eq6.5}
 \ee
 It follows, that if $x$ is close to one, $\beta_1$ is approximately $1/2$.
Since $\beta_1 = 1/2$ is a non-singular point, computation of master integrals
simplifies.

 The $\theta \delta$ and $\theta \theta$ integrals are similar. We consider the
latter case, as it is more general. We assume that $\alpha_1 = \beta_1/z_1$ and
$\beta_2 = \alpha_2/z_2$. Using this parameterization, we find the following
expression for $2 k_1 k_2$
 \be
 2 k_1 k_2 = \frac{ \beta_1 \alpha_2 }{z_1 z_2}  \left ( 1 - 2 \sqrt{z_1 z_2} \cos \varphi_{12} + z_1 z_2 \right ).
 \ee
 This expression simplifies if we consider the limit where both $z_1$ and $z_2$
become vanishingly small since the dependence on the azimuthal angle disappears.
This simplification allows us to compute the boundary conditions for master
integrals with a relative ease. We also note that a similar argument applies to
$\theta \delta$-integrals since in that case we obtain $2 k_1 k_2$ from the
above equation by setting either $z_1$ or $z_2$ to 1.

 We will now discuss two examples of the boundary integrals. One of the $\delta
\delta$-integrals that we need reads
 \be
J(x) = \int {\rm d} \Phi^{nn}_{\delta \delta} \; \delta(2 k_1 k_2 -x ) \; \pent{2,1}{11101}, 
\ee
where the one-loop integral can be found in Appendix~\ref{app:1}. Using the
expression for $\pent{2,1}{11101}$, and the fact that ${\rm d} \Phi^{nn}_{\delta
\delta} \delta(2 k_1 k_2 - x)$ equals to ${\rm d} \Phi$ in Eq.~(\ref{eq6.1}), we
find
\be
\begin{split} 
J(x) & =i \left (  \frac{ \Omega^{(d-2)}}{4 (2\pi)^{d-1}} \right )^3 e^{-i \pi \ep}
\left ( \frac{\Gamma(1-\ep) \Gamma(1+\ep)}{\ep} \right )^2 \; N_x \; x^{-1/2 -\ep}
\\
& \times \int {\rm d} \beta \left (  4 \beta (1-\beta) - x \right )^{-1/2-\ep} (  1-\beta )^{-1-\ep}
  {}_2F_1 \left ( 1 + \ep, -\ep, 1-\ep; 1 - \beta \right ),
\end{split} 
\ee
where the integration over $\beta$ is performed on the interval shown in
Eq.~(\ref{eq6.5}).

We then consider the limit $x \to 1$ in which case, as we explained earlier, the
integration over $\beta$ is restricted to $\beta \approx 1/2$.  It follows that
\be
\begin{split} 
& \lim_{x \to 1} I(x)  \approx i \left (  \frac{ \Omega^{(d-2)}}{4 (2\pi)^{d-1}} \right )^3 e^{-i \pi \ep}
\left ( \frac{\Gamma(1-\ep) \Gamma(1+\ep)}{\ep} \right )^2 \; N_x \;
\\
& \times 2^{1+\ep} {}_2F_1 \left ( 1 + \ep, -\ep, 1-\ep; 1/2 \right )
\; \frac{ \Gamma(1/2) \Gamma(1/2-\ep) }{ 2 \Gamma(1-\ep) } \; (1-x)^{-\ep},
\end{split}
\ee
where we used  
\be
\lim_{x \to 1} \; 
\int 
\limits_{
 \frac{1}{2} ( 1 - \sqrt{1-x} ) }^{ \frac{1}{2} ( 1 + \sqrt{1 - x} )}
\;
{\rm d} \beta \left (  4 \beta (1-\beta) - x \right )^{-1/2-\ep}
\approx  \frac{\Gamma(1/2) \Gamma(1/2-\ep)}{ 2 \Gamma(1-\ep)} \; (1-x)^{-\ep}.
\ee
Finally, we note that all  other $x$-dependent integrals can be studied in a similar fashion. 
\\

We continue with the discussion of the boundary conditions for $\theta \delta$-
and $\theta \theta$-integrals. We will only consider one of the $\theta \theta$
integrals since the analysis of the $\theta \delta$ ones is rather similar. To
this end, consider the following integral
\be
J_{\theta \theta}(z_1,z_2)  = \int [{\rm d} k_1 ][{\rm d} k_2]
\delta(1 - \beta_1 - \alpha_2) \delta( z_1 \alpha_1 - \beta_1) \delta(z_2 \beta_2 - \alpha_2)
\; \frac{\pent{2,1}{11101}}{ 2 k_1 \cdot k_2 \; k_{12} \cdot n}.
\ee
Using the expression for $\pent{2,1}{11101}$ from Appendix~\ref{app:1}, we write 
\be
\begin{split} 
  & J_{\theta \theta}(z_1,z_2)  = i \left ( \frac{ \Omega^{(d-2)}}{4 (2 \pi)^{d-1} } \right )^3
    e^{-i \pi \ep} \left [\frac{\Gamma(1-\ep) \Gamma(1+\ep)}{\ep} \right ]^2
    \int \prod \limits_{i=1}^{2} {\rm d} \alpha_i {\rm d} \beta_i (\alpha_i \beta_i)^{-\ep}
    \frac { {\rm d} \Omega_{\varphi_{12}}^{(d-2)} }{\Omega^{(d-2)}} 
  \\
  & \times 
    \delta(1 - \beta_1 - \alpha_2) \delta( z_1 \alpha_1 - \beta_1) \delta(z_2 \beta_2 - \alpha_2) \;
    \frac{z_1 z_2 }{\beta_{12} \beta_1 \alpha_2 (1-2 \sqrt{z_1 z_2} \cos \varphi_{12} + z_1 z_2)}
  \\
  & \times ( \beta_2 \alpha_{12} )^{-1-\ep} {}_2F_1 \left ( 1+\ep, -\ep, 1-\ep; \frac{\alpha_2}{\alpha_{12}} \right ).
\end{split} 
\ee
We observe that at small values of $z_{1,2}$ the dependence on $\cos
\varphi_{12}$ disappears from the integrand. Hence, the leading asymptotic
behavior of $J_{\theta \theta}(z_1,z_2)$ in the limit $z_{1,2} \to 0$, follows
from the simplified integral
\be
\begin{split} 
  & \lim_{z_{1,2} \to 0}^{} J_{\theta \theta} \sim
    i \left ( \frac{ \Omega^{(d-2)}}{4 (2 \pi)^{d-1} } \right )^3
    e^{-i \ep \pi} \left [\frac{\Gamma(1-\ep) \Gamma(1+\ep)}{\ep} \right ]^2
    z_1^{1+2\ep} z_2^{2 + 2\ep}\;
    F_{\theta \theta}(z_1,z_2),
  \\
 & F_{\theta \theta} = \int \limits_{0}^{1} 
   \frac{
   {\rm d} \beta_1 \;
   (1-\beta_1)^{-1-2 \ep} \beta_1^{-2 -3 \ep}  
   }{ ( 1- \beta_1 + \beta_1 z_1)^{1+\ep}
   ( \beta_1 + (1-\beta_1) z_2) } \; {}_2F_1\left(1+\ep,-\ep,1-\ep;\frac{\beta_1 z_1}{1-\beta_1 +
   \beta_1 z_1}\right).
\end{split} 
\ee
To integrate over $\beta_1$ in the limit $z_{1,2} \to 0$, we note that three
integration region lead to particular dependencies on $z_{1,2}$; they are
$\beta_1 \sim z_2$, $ z_2 \ll \beta_1 \ll 1-z_1$ and $ 1- z_1 \sim \beta_1 $.
Contributions of each of these regions can be calculated independently of each
other by expanding the integral in variables that are small in a particular
region and extending the integration region to ensure that homogeneous scaling
of the remaining integral in the small variable is achieved
\cite{Beneke:1997zp}. Since we are interested in the leading asymptotic in each
region, this is a relatively simple thing to do. Hence, we write
\be
\lim_{z_{1,2} \to 0}^{} F_{\theta \theta}(z_1,z_2) =
  F_{T} + F_{\beta_1 \sim z_2} + F_{\beta_1 \sim 1-z_1}. 
\ee

The contribution of the ``Taylor-expansion'' region $F_T$ is computed by setting
$z_1 = z_2 =0$ everywhere in the integrand of the function $F_{\theta \theta} $.
We then find
\be
F_T = \int \limits_{0}^{1} {\rm d} \beta_1
(1-\beta_1)^{-2-3 \ep} \beta_1^{-3 -3 \ep} =
\frac{\Gamma(-2-3\ep) \Gamma(-1-3\ep)}{\Gamma(-3 - 6 \ep)}.
\ee
In the region where $\beta_1 \sim z_2$ the argument of the hypergeometric
function is small, ${\cal O}(z_2 z_1)$, so that the hypergeometric function can
be Taylor-expanded. Furthermore, to leading order in $\beta_1 \sim z_2 \ll 1$,
we can replace
\be
1-\beta_1 + \beta_1 z_1 \to 1, \;\;\; 1-\beta_1
\to 1, \;\;\; \beta_1 + (1-\beta_1) z_2 \to \beta_1 + z_2.
\ee
To ensure that the resulting integral scales uniformly with $z_2$, we extend the
integration over $\beta_1$ to infinity and find
\be
F_{\beta_1 \sim z_2} = \int \limits_{0}^{\infty} {\rm d} \beta_1 \;
\beta_1^{-2-3\ep} ( \beta_1 + z_2)^{-1} = z_2^{-2-3\ep} \; \Gamma(-1-3\ep) \Gamma(2+3\ep).
\ee
Finally, we perform a similar analysis for $\beta_1 \sim 1-z_1$. The only
difference with respect to the previous case is that the hypergeometric function
cannot be simplified since $\beta_1 z_1/(1-\beta_1 + \beta_1 z_1) \sim O(1)$ in
this region. Hence, changing variables $\beta_1 = 1-\xi$, we write
\be
F_{\beta_1 \sim 1- z_1} =
\int \limits_{0}^{\infty} {\rm d} \xi \; \frac{ \xi^{-1-2\ep}}{(\xi + z_1)^{1+\ep}} {}_2F_1 \left ( 1+ \ep, -\ep,1-\ep;\frac{z_1}{\xi+z_1} \right ). 
\ee
Writing $\xi  = z_1 (1-u)/u$, we find
\be
\begin{split} 
  & F_{\beta_1 \sim 1- z_1} = z_1^{-1-3\ep} \int \limits_{0}^{\infty} {\rm d} u \; u^{3\ep} (1-u)^{-1-2\ep}
  {}_2F_1 \left ( 1+ \ep, -\ep,1-\ep;u \right )
  \\
  & =
    z_1^{-1-3\ep} \; \frac{\Gamma(1+3\ep) \Gamma(-2\ep)}{1+\ep} {}_3F_2( \{1+\ep,-\ep,1+3\ep \},\{1-\ep,1+\ep \};1).
\end{split} 
\ee
   We note that each of the computed contributions is the leading contribution
for a particular ``branch'' since each of them involves a dependence on
$z_{1,2}$ raised to a particular $\ep$-dependent power and each receives ${\cal
O}(z_1,z_2)$ corrections. Some of the computed branches can be used as the
boundary conditions for the corresponding integral whereas other follow from the
inhomogeneous (simpler) terms in the corresponding differential equation and can
be used for consistency checks.

The approach described above drastically simplifies the calculation of the
required boundary conditions for all integrals except the pentagons, see
Eqs.~(\ref{eqb.6}) and (\ref{eqb.61}). However, it follows from the differential
equations constructed for these master integrals and from the analysis of their
Feynman-parameter representation that no independent boundary conditions are
required for integrals of the five-point functions and the results for them can
be obtained by simply integrating the inhomogeneous terms in the corresponding
differential equations.

\section{Results}
\label{seq7}

Following the above discussion, we obtain the results for the gluon, ghost and
quark-antiquark contributions to the soft function. We emphasize that the
separation of final states into ``gluons'' and ``ghosts'' is unphysical; it is
related to the fact that we use $-g^{\mu \nu}$ to describe the density matrix of
the final-state gluons. Hence, we only present the result for the correct
combination of gluons and ghosts, and quarks below.\footnote{Separate results
for the ghost contribution as well as for the \emph{unphysical} Feynman-gauge
gluon contribution can be found in an ancillary file provided with this
submission.}

To present the one-loop correction to the double-real emission contribution to
the zero-jettiness soft function, we write
\begin{equation}
  \label{eq:resSplit}
  S_{RRV}^{(3)} = \tau^{-1-6\ep}
  \cos\left( \pi \varepsilon \right)\left ( 
    \frac{g^2_s}{16 \pi^2} \frac{(4\pi)^\ep}{\Gamma(1-\ep)}
  \right )^3
  \left ( S_{\textrm{RRV},gg}^{(3)} + S_{\textrm{RRV},q \bar q}^{(3)}\right ),  
\end{equation}
 where $g_s$ is the bare QCD coupling constant. We note that we need to compute
$S_{\textrm{RRV},gg}^{(3)}$ and $ S_{\textrm{RRV},q \bar q}^{(3)}$ through
${\cal O}(\ep)$ since the $\ep$-expansion of the factor $\tau^{-1-6\ep}$ that
appears in Eq.~(\ref{eq:resSplit}) generates a single $1/\ep$ pole
\be
\tau^{-1-6 \ep} = -\frac{\delta(\tau) }{6 \ep} + 
\cdots . 
\ee

  Physically, hard emitters can be either in the fundamental or in the adjoint
representation of QCD; these cases describe processes with two hard quarks or
gluons, respectively. Below we present the result of the calculation of the
one-loop correction to the emission of two partons where color charges of hard
emitters are not specified. We find
  \begin{align}
    \label{eq:resSgg}
    S_{\textrm{RRV},gg}^{(3)}
    & = 
      \cR \cA^2 
      \Biggl[
      -\frac{40}{3 \ep^5}
      -\frac{88}{3 \ep^4}
      -\frac{1}{\ep^3} \left ( 
      \frac{2144}{27}
      -\frac{76 \pi^2}{9}
    \right ) 
      -\frac{1}{\ep^2} \left ( 
      \frac{16448}{81}
      -\frac{220 \pi ^2}{27}
      -336 \; \zeta_3      
  \right )
    \nonumber\\
    & 
      -\frac{1}{\ep}
      \left ( \frac{33832}{81}
      +\frac{17152 \pi ^2}{81}
      -\frac{14432\zeta_3}{9}
      -\frac{2677 \pi ^4}{135}
     \right )
      -\frac{641296}{729}
      +\frac{12880 \pi ^2}{243}
    \nonumber\\
    & 
          -\frac{81472 \zeta_3}{9}
      +\frac{14311 \pi ^4}{135}
      -\frac{872 \pi ^2 \zeta_3}{9}
      +\frac{18880 \zeta_5}{3}
      -\ep \Biggl(
      \frac{2365264}{2187}
      +\frac{896476 \pi^2}{729}
\nonumber\\
    & 
          -\frac{724720 \zeta_3}{81}
      +\frac{51188 \pi^4}{81}
      +\frac{121616 \pi^2 \zeta_3}{27}
      -\frac{325600 \zeta_5}{3}
      -\frac{229231 \pi^6}{17010}
 \nonumber\\
     &  
      -\frac{17440 \zeta_3^2}{3}
      \Biggr)      
      \Biggr]
     + \cR\cA  \left(\nf\TF\right)
      \Biggl[
      \frac{16}{9\ep^2}
      +\frac{400}{27 \ep}
      +\frac{8 \pi^2}{9}
      +\frac{1568}{27}
      \nonumber \\
      & 
      +\ep \left(     
      \frac{44432}{243}
      +\frac{56 \pi ^2}{3}
      -\frac{736 \zeta_3}{9}
      \right)
      \Biggr]
       +\cR^2 \cA
      \Biggl[ 
      \frac{48}{\ep^5}
      -\frac{56 \pi ^2}{\ep^3}
      -\frac{2400 \zeta_3}{\ep^2}
\nonumber \\
     &  
      -\frac{526 \pi ^4}{5 \ep}
      -64800 \zeta_5
      +2800 \pi ^2 \zeta_3
       +\ep \left(60000 \zeta_3^2-\frac{22271 \pi ^6}{105}\right)\Biggr] + \mathcal{O}\left( \ep^2 \right),
\end{align}
and 
\begin{align}
  \label{eq:resSqq}
  S_{\textrm{RRV}, q\bar{q}}^{(3)}
  & = \cR \; \cA \; ( \nf\TF  ) \;  
  \Biggl[  
    \frac{32}{9 \ep^4}
    +\frac{16}{27 \ep^3}
    -\frac{1}{\ep^2}\left(
    \frac{544}{81}
    -\frac{16 \pi^2}{27}    
    \right)
    -\frac{1}{\ep}
    \left(\frac{2096}{243}
    -\frac{296 \pi ^2}{81}\right)
    \nonumber\\
  & 
    +\frac{139808}{729}
    +\frac{7360 \pi ^2}{243}
    -256 \zeta_3
    -\frac{92 \pi^4}{135}
    +\ep \Bigl(
    \frac{1634224}{2187}
    -\frac{382360 \pi^2}{729}
    +\frac{43328 \zeta_3}{9}
    \nonumber\\
  & 
    -\frac{5182 \pi^4}{405}
    -\frac{2816 \pi^2 \zeta_3}{27}
    +\frac{4864 \zeta_5}{3}
    \Bigr)
    \Biggr]
    + \cR \; \left( \nf\TF \right)^2
    \Biggl[ 
    \frac{128}{27 \ep^3}
    +\frac{1280}{81\ep^2}
    +\frac{1}{\ep}\left(\frac{128}{3}-\frac{64 \pi^2}{27}\right)
    \nonumber\\
    & 
      -\frac{5120}{729}
      +\frac{4352 \pi ^2}{81}
      -\frac{10496 \zeta_3}{27}
      -\ep \Bigl(
      \frac{839552}{2187}
      +\frac{1088 \pi ^2}{27}
      -\frac{224512 \zeta_3}{81}
      +\frac{1136 \pi ^4}{45}
      \Bigr)
      \Biggr]
      \nonumber\\
  & + \cR \; \CF \; ( \nf\TF )
    \Biggl[ 
    \frac{64}{9 \ep^4}
    +\frac{608}{27 \ep^3}
    +\frac{1}{\ep^2}\left( \frac{5536}{81}-\frac{32 \pi ^2}{9} \right)
    +\frac{1}{\ep}\Bigl(
    \frac{2432}{243}
    +\frac{2192 \pi^2}{27}
    -\frac{5248 \zeta_3}{9}
    \Bigr)
  \nonumber\\
  & 
    -\frac{352832}{729}
    -\frac{6320 \pi^2}{81}
    +\frac{114880 \zeta_3}{27}
    -\frac{568 \pi^4}{15}
    - \ep \Bigl(
    \frac{7722208}{2187}
    -\frac{249632 \pi^2}{243}
    +\frac{812800 \zeta_3}{81}
    \nonumber\\
  & 
    -\frac{7828 \pi^4}{27}
    -\frac{15680 \pi^2 \zeta_3}{9}
    + 41088 \zeta_5
    \Bigr)
    \Biggr] + \mathcal{O}\left( \ep^2 \right).
\end{align}

The label $R = F,A$ in the above formula refers to the $SU(3)$ representation of
the hard emitters; furthermore, $\CF=(\NC^2-1)/2\NC$, $\cA=\NC$ and $T_F=1/2$
are the usual Casimir invariants of the gauge group $SU(N_c)$.

We have checked the above result in several ways. First, we observe that there
is no ${\cal O}(C_R^3)$ contribution in $S^{(3)}_{RRV,gg}$ and $C_R^2 n_f T_F$
contribution in $S^{(3)}_{RRV, q \bar q}$. This is the consequence of the
exponentiation of soft emissions in Abelian gauge theories as follows from
considering $R = F$ case. Next, the complete input for computing tree- and
one-loop eikonal currents was assembled using two different methods. All master
integrals as well as some integrals that appear in the soft function before the
reduction to master integrals were computed numerically and compared with the
analytic results. For the numerical evaluation, we used
\texttt{pySecDec}~\cite{Heinrich:2023til} and \texttt{MB}~\cite{Czakon:2005rk}
packages.

Finally, our results for $gg$ and $c \bar c$ final states agree with their
recent evaluation reported in Ref.~\cite{Chen:2020dpk}. We note that the result
for $S_{RRV}^{gg}$ reported in the ancillary file of Ref.~\cite{Chen:2020dpk}
appears to contain irrational numbers of weight seven which are absent in our
result. However, all these weight seven quantities disappear once the result
reported in Ref.~\cite{Chen:2020dpk} is carefully simplified.

\section{Conclusions}
\label{seq8}

In this article, we described an analytic computation of the one-loop correction
to the double-real emission contribution to the zero-jettiness soft function.
Our calculation is based on reverse unitarity and its modification
\cite{Baranowski:2021gxe} required to deal with phase-space integrals that
contain Heaviside functions. Our result for $gg$ final state agrees with the
recent computation of the same quantity reported in Ref.~\cite{Chen:2020dpk}. In
addition, we also presented the one-loop correction to the contribution of the
$q \bar q$ final state to the zero-jettiness soft function.

We explained in detail the way to construct the input for tree- and one-loop
two-parton currents and the organization of the integration-by-parts reduction
since publicly available tools cannot be used for this purpose when Heaviside
functions are present. We discussed how master integrals are calculated; we
present results for all of them in an ancillary file provided with this
submission.

Finally, we briefly comment on what remains to be done to complete the analytic
calculation of the zero-jettiness soft function. The most non-trivial
contribution that still needs to be calculated is the part of the triple-real
emission contribution that describes the kinematic configuration where two
gluons are radiated into one hemisphere and the third gluon -- into the opposite
hemisphere. We are quite confident that methods employed in
Ref.~\cite{Baranowski:2022khd} to compute the same-hemisphere triple-real
contribution to the soft function can be successfully adapted to this case as
well, and we look forward to addressing this challenge.
\\

\acknowledgments
We are grateful to Rayan~Haindl for participating in the initial stages of this project.
The research of KM and AP was partially supported by the Deutsche Forschungsgemeinschaft
(DFG, German Research Foundation) under grant 396021762-TRR 257.
The research of MD was supported by the European Research Council (ERC) 
under the European Union’s research and innovation programme grant agreement 949279 (ERC Starting Grant HighPHun).
The work of DB is supported in part by the Swiss National Science Foundation (SNSF) under contracts 200020$\mathrm{\_}$188464 and 200020$\mathrm{\_}$219367.

\appendix

\section{One-loop integrals}
\label{app:1}

\begin{equation}
  \label{eq:pent-fig}
  \pent{i,j}{a_1a_2a_3a_4a_5}  = \vcenter{\hbox{
      \begin{tikzpicture}
        \draw[WLBE] (0,0) -- node[pos=0.88, anchor = south east] {$n$} (2,2)  {};
        \draw[WLBS] (2,-2) -- node[pos=0.12, anchor = north east] {$\bar{n}$} (0,0)  {};
        \draw[-{Stealth[scale=0.8]}] (1.5,1.7)  -- (1.8,2.0)  {};
        \draw[-{Stealth[scale=0.8]}] (1.5,-1.7)  -- (1.8,-2.0)  {};
        \draw (1.5,1.5) -- (2,0.6);
        \draw (1.5,-1.5) -- (2,-0.6);
        \draw (2,0.6) -- (2,-0.6);
        \draw (2,0.6) -- (2.5,0.8);
        \draw (2,-0.6) -- (2.5,-0.8);
        \node[anchor=south east] at (0.75,0.75) {\scriptsize$a_1$};
        \node[anchor=west] at (1.8,1.15) {\scriptsize$a_2$};
        \node[anchor=west] at (2,0) {\scriptsize$a_3$};
        \node[anchor=west] at (1.8,-1.15) {\scriptsize$a_4$};
        \node[anchor=north east] at (0.75,-0.75) {\scriptsize$a_5$};
        \node[anchor=west] at (2.5,0.8) {$k_i$};
        \node[anchor=west] at (2.5,-0.8) {$k_j$};
        \fill (1.5,1.5) circle (1pt);
        \fill (1.5,-1.5) circle (1pt);
        \fill (2,0.6) circle (1pt);
        \fill (2,-0.6) circle (1pt);
        \fill (-0.1,-0.1) rectangle (0.1,0.1);
        \path[use as bounding box] (-0.3,-2.5) rectangle (3,2.5);
      \end{tikzpicture}
    }},
  \quad
    \pbox{i}{a_1a_2a_3a_4}  = \vcenter{\hbox{
      \begin{tikzpicture}
        \draw[WLBE] (0,0) -- node[pos=0.88, anchor = south east] {$n$} (2,2)  {};
        \draw[WLBS] (2,-2) -- node[pos=0.12, anchor = north east] {$\bar{n}$} (0,0)  {};
        \draw[-{Stealth[scale=0.8]}] (1.5,1.7)  -- (1.8,2.0)  {};
        \draw[-{Stealth[scale=0.8]}] (1.5,-1.7)  -- (1.8,-2.0)  {};
        \draw (1.5,1.5) -- (2,0);
        \draw (1.5,-1.5) -- (2,0);
        % %
        \draw (2,0) -- (2.5,0);
        \node[anchor=south east] at (0.75,0.75) {\scriptsize$a_1$};
        \node[anchor=west] at (1.8,0.9) {\scriptsize$a_2$};
        \node[anchor=west] at (1.8,-0.9) {\scriptsize$a_3$};
        \node[anchor=north east] at (0.75,-0.75) {\scriptsize$a_4$};
        \node[anchor=west] at (2.5,0) {$k_i$};
        \fill (1.5,1.5) circle (1pt);
        \fill (1.5,-1.5) circle (1pt);
        \fill (2,0) circle (1pt);
        \fill (-0.1,-0.1) rectangle (0.1,0.1);
        \path[use as bounding box] (-0.3,-2.5) rectangle (3,2.5);
      \end{tikzpicture}
    }}  
\end{equation}

\begin{align}
  \label{eq:pent-box-props}
  \pent{i,j}{a_1a_2a_3a_4a_5} & = \int \frac{\dm^d l}{(2\pi)^d} \frac{1}{
                                \left[ l\cdot n \right]^{a_1}
                                \left[ l^2 \right]^{a_2}
                                \left[ (l-k_i)^2 \right]^{a_3}
                                \left[ (l-k_{ij})^2 \right]^{a_4}
                                \left[ (k_{ij}-l)\cdot \bar{n} \right]^{a_5}} \,,\\
  \pbox{i}{a_1a_2a_3a_4} & = \int \frac{\dm^d l}{(2\pi)^d} \frac{1} {
                           \left[  l\cdot n \right]^{a_1}
                           \left[  l^2\right]^{a_2}
                           \left[  (l-k_i)^2  \right]^{a_3}
                           \left[  (k_i-l)\cdot \bar{n} \right]^{a_4}} \,,
\end{align}
where $k_{ij}=k_i+k_j$.

For the calculations described in
this paper, we need a number  of loop integrals. The results
that we have relied upon are given below. 
\begin{align} 
\allowdisplaybreaks
  % L_1
  \pent{1,2}{01010} & =  \int \frac{{\rm d}^d l}{(2 \pi)^d} \; \frac{1}{ [ l^2 ]  \;\; [ (l - k_{12})^2 ]} =  \frac{i}{(4 \pi)^{d/2}} \; \frac{\Gamma^2(1-\ep) \Gamma(\ep)}{\Gamma(2-2\ep)}\; e^{-i \pi \ep} \left ( k^2_{12} \right )^{-\ep},
  \\  \nonumber  \\
  % L_2
  \pent{2,1}{10101} & = \int \frac{{\rm d}^d l}{(2 \pi)^d} \; \frac{1}{ [ l \cdot n ] \;  [(l - k_2)^2 ] \; [ (k_{12} - l ) \cdot \bar n ] }
  \\
                    & = i \frac{\Omega^{(d-2)}}{4 (2\pi)^{d-1}}  e^{-i \pi \ep}
                      \left[ \frac{ \Gamma(1-\ep) \Gamma(1+\ep)}{\ep}  \right ]^2 \; ( \alpha_1 \beta_2 )^{-\ep},
 \nonumber \\ \nonumber \\
  % L_3
  \pent{1,2}{11011} & = \int \frac{{\rm d}^d l}{(2 \pi)^d} \; \frac{1}{ [ l \cdot n ]  \; [ l^2 ]  [ (l - k_{12})^2 ]
                      [ (k_{12} - l ) \cdot \bar n ] } =  \frac{i \Omega^{(d-2)}}{4 (2\pi)^{d-1}}  e^{-i \pi \ep}\;
  \\
                    &   \times \frac{\Gamma^3(1-\ep) \Gamma^2(1+\ep)}{
                      \ep^2 \Gamma(-2\ep) \Gamma(2+\ep) \; (\alpha_{12} \beta_{12} )^{1+\ep} } {}_2F_1\left ( 1+\ep,1+\ep,2+\ep;1 - \frac{k_{12}^2}{\alpha_{12} \beta_{12} }
                      \right ),
                      \nonumber \\ \nonumber \\
  % L_4
  \pent{2,1}{11110} & =  \int \frac{{\rm d}^d l}{(2 \pi)^d} \;  \frac{1}{ [ l \cdot n ] \; [ l^2 ] [ (l-k_2)^2 ] [ (l - k_{12})^2 ]}
\\
                    & = \frac{i \Gamma(1-\ep) \Gamma(-\ep) \Gamma(\ep)}{ (4\pi)^{d/2} \ep \Gamma(-2\ep) } e^{-i \pi \ep}
                      \frac{ \left ( k_{12}^2 \right )^{-1-\ep} }{\beta_2^{1+\ep} } ( \beta_{12} )^\ep {}_2F_1 \left ( -\ep,-\ep,1-\ep;
                      \frac{\beta_1}{\beta_{12}} \right ),
                      \nonumber \\  \nonumber \\
  \pent{2,1}{10111}   & = \int \frac{{\rm d}^d l}{(2 \pi)^d} \; \frac{1}{ [  l \cdot n ]  \; [  (l - k_2)^2 ] [  (l - k_{12})^2 ] [  (k_{12} - l ) \cdot \bar n ] }
  \\
                    & = 
                      i \frac{\Omega^{(d-2)}}{4 (2\pi)^{d-1}}  e^{-i \pi \ep}  \left[ \frac{ \Gamma(1-\ep) \Gamma(1+\ep)}{\ep} \right ]^2
                      ( \alpha_1 \beta_{12} )^{-1-\ep} {}_2F_1\left ( 1+\ep,-\ep,1-\ep;\frac{\beta_1}{\beta_{12}} \right ),
                      \nonumber \\ \nonumber \\
  % L_6
  \pbox{1}{1111} & = \int \frac{{\rm d}^d l}{(2 \pi)^d} \frac{1}{[ l \cdot n] \; [ l^2 ] \;  [(l - k_1)^2]  [ (k_{1} - l ) \cdot \bar n ]}
  \\
                    & = 
                      i \frac{\Omega^{(d-2)}}{4 (2\pi)^{d-1}}  e^{-i \pi \ep}  \left[ \frac{ \Gamma(1-\ep) \Gamma(1+\ep)}{\ep} \right ]^2
                      ( \beta_1 \alpha_1 )^{-1-\ep} \frac{\Gamma(-\ep) \Gamma(1-\ep)}{\Gamma(-2\ep)}\,,
\nonumber 
\end{align}
where $\Omega^{(n)}=2\pi^{n/2}/\Gamma(n/2)$.

\section{List of master integrals}
\label{sec:mi-list}

To present a list of master integrals, we define
\begin{align}
  & \small{\texttt{II[C,P,\{F,n1,...,nX\},a,\{b1,b2,b3\},\{c1,c2,c3\}]}} = \nonumber \\
  & \quad \int \frac{\dm\Phi^{\omega(C) }_{P} \; \textrm{F}_{n_1\dots n_X}}{
    (k_1 \cdot k_2)^a
    (k_1 \cdot n)^{b_1}
    (k_2 \cdot n)^{b_2}
    (k_{12} \cdot n )^{b_3}
    (k_1 \cdot \bar{n})^{c_1}
    (k_2 \cdot \bar{n})^{c_2}
    (k_{12} \cdot \bar{n} )^{c_3}
    },\nonumber
\end{align}
where we will use the notation $\omega(A) = n n$ and $\omega(B) = n \bar n$
below. The complete list of master integrals reads
\allowdisplaybreaks
\begin{alignat}{2}
  \label{eq:II-mis}
  & \texttt{\small{II[A,dd,\{B1,1,1,1,1\},0,\{0,0,0\},\{0,0,0\}]}} && = \int \dFddA\pbox{1}{1111}, \\
  & \texttt{\small{II[A,dd,\{P12,0,1,0,1,0\},0,\{0,0,0\},\{0,0,0\}]}} && = \int \dFddA\pent{1,2}{01010}, \\
  & \texttt{\small{II[A,dd,\{P12,0,1,1,1,1\},0,\{0,0,0\},\{0,0,0\}]}} && = \int \dFddA\pent{1,2}{01111}, \\
  & \texttt{\small{II[A,dd,\{P12,1,1,0,1,1\},0,\{0,0,0\},\{0,0,0\}]}} && = \int \dFddA\pent{1,2}{11011}, \\
  & \texttt{\small{II[A,dd,\{P12,1,1,0,1,1\},0,\{0,0,0\},\{1,0,0\}]}} && = \int  \frac{\dFddA\pent{1,2}{11011}}{(k_1 \cdot \bar{n})}, \\
  & \texttt{\small{II[A,dd,\{P12,1,1,1,1,1\},0,\{0,0,0\},\{0,0,0\}]}} && = \int \dFddA\pent{1,2}{11111}, 
                                                                         \label{eqb.6} \\
  & \texttt{\small{II[A,dd,\{P21,1,0,1,0,1\},0,\{0,0,0\},\{0,0,0\}]}} && = \int \dFddA\pent{2,1}{10101}, \\
  & \texttt{\small{II[A,dd,\{P21,1,0,1,1,1\},0,\{0,0,0\},\{0,0,0\}]}} && = \int \dFddA\pent{2,1}{10111}, \\
  & \texttt{\small{II[A,td,\{P12,0,1,0,1,0\},0,\{0,0,0\},\{0,0,1\}]}} && = \int  \frac{\dFtdA\pent{1,2}{01010}}{(k_{12} \cdot \bar{n} )}, \\
  & \texttt{\small{II[A,td,\{P12,0,1,0,1,0\},0,\{0,0,0\},\{1,0,0\}]}} && = \int  \frac{\dFtdA\pent{1,2}{01010}}{(k_1 \cdot \bar{n})}, \\
  & \texttt{\small{II[A,td,\{P12,0,1,0,1,0\},0,\{1,0,0\},\{0,0,1\}]}} && = \int  \frac{\dFtdA\pent{1,2}{01010}}{(k_1 \cdot n)(k_{12} \cdot \bar{n} )}, \\
  & \texttt{\small{II[A,td,\{P12,0,1,0,1,0\},1,\{0,1,1\},\{0,0,1\}]}} && = \int  \frac{\dFtdA\pent{1,2}{01010}}{(k_1 \cdot k_2)(k_2 \cdot n)(k_{12} \cdot n)(k_{12} \cdot \bar{n} )}, \\
  & \texttt{\small{II[A,td,\{P12,0,1,0,1,0\},1,\{1,0,1\},\{0,0,1\}]}} && = \int  \frac{\dFtdA\pent{1,2}{01010}}{(k_1 \cdot k_2)(k_1 \cdot n)(k_{12} \cdot n)(k_{12} \cdot \bar{n} )}, \\
  & \texttt{\small{II[A,td,\{P12,0,1,0,1,0\},1,\{1,1,0\},\{0,0,1\}]}} && = \int  \frac{\dFtdA\pent{1,2}{01010}}{(k_1 \cdot k_2)(k_1 \cdot n)(k_2 \cdot n)(k_{12} \cdot \bar{n} )}, \\
  & \texttt{\small{II[A,td,\{P12,0,1,1,1,1\},0,\{0,0,0\},\{0,0,1\}]}} && = \int  \frac{\dFtdA\pent{1,2}{01111}}{(k_{12} \cdot \bar{n} )}, \\
  & \texttt{\small{II[A,td,\{P12,0,1,1,1,1\},0,\{1,0,0\},\{0,0,1\}]}} && = \int  \frac{\dFtdA\pent{1,2}{01111}}{(k_1 \cdot n)(k_{12} \cdot \bar{n} )}, \\
  & \texttt{\small{II[A,td,\{P12,1,0,1,0,1\},0,\{0,0,0\},\{0,0,1\}]}} && = \int  \frac{\dFtdA\pent{1,2}{10101}}{(k_{12} \cdot \bar{n} )}, \\
  & \texttt{\small{II[A,td,\{P12,1,0,1,0,1\},1,\{0,0,0\},\{0,0,1\}]}} && = \int  \frac{\dFtdA\pent{1,2}{10101}}{(k_1 \cdot k_2)(k_{12} \cdot \bar{n} )}, \\
  & \texttt{\small{II[A,td,\{P12,1,0,1,0,1\},1,\{1,0,0\},\{0,1,1\}]}} && = \int  \frac{\dFtdA\pent{1,2}{10101}}{(k_1 \cdot k_2)(k_1 \cdot n)(k_2 \cdot \bar{n})(k_{12} \cdot \bar{n} )}, \\
  & \texttt{\small{II[A,td,\{P12,1,0,1,0,1\},1,\{1,0,0\},\{1,0,1\}]}} && = \int  \frac{\dFtdA\pent{1,2}{10101}}{(k_1 \cdot k_2)(k_1 \cdot n)(k_1 \cdot \bar{n})(k_{12} \cdot \bar{n} )}, \\
  & \texttt{\small{II[A,td,\{P12,1,0,1,0,1\},1,\{1,0,0\},\{1,1,0\}]}} && = \int  \frac{\dFtdA\pent{1,2}{10101}}{(k_1 \cdot k_2)(k_1 \cdot n)(k_1 \cdot \bar{n})(k_2 \cdot \bar{n})}, \\
  & \texttt{\small{II[A,td,\{P12,1,0,1,0,1\},1,\{2,0,0\},\{0,0,1\}]}} && = \int  \frac{\dFtdA\pent{1,2}{10101}}{(k_1 \cdot k_2)(k_{12} \cdot \bar{n} )}, \\
  & \texttt{\small{II[A,td,\{P12,1,0,1,1,1\},0,\{0,0,0\},\{0,0,1\}]}} && = \int  \frac{\dFtdA\pent{1,2}{10111}}{(k_{12} \cdot \bar{n} )}, \\
  & \texttt{\small{II[A,td,\{P12,1,0,1,1,1\},1,\{1,0,0\},\{0,0,1\}]}} && = \int  \frac{\dFtdA\pent{1,2}{10111}}{(k_1 \cdot k_2)(k_1 \cdot n)(k_{12} \cdot \bar{n} )}, \\
  & \texttt{\small{II[A,td,\{P12,1,1,0,1,1\},0,\{0,0,0\},\{0,0,0\}]}} && = \int \dFtdA\pent{1,2}{11011}, \\
  & \texttt{\small{II[A,td,\{P12,1,1,0,1,1\},0,\{0,0,0\},\{0,0,1\}]}} && = \int  \frac{\dFtdA\pent{1,2}{11011}}{(k_{12} \cdot \bar{n} )}, \\
  & \texttt{\small{II[A,td,\{P12,1,1,0,1,1\},1,\{0,0,0\},\{1,0,0\}]}} && = \int  \frac{\dFtdA\pent{1,2}{11011}}{(k_1 \cdot k_2)(k_1 \cdot \bar{n})}, \\
  & \texttt{\small{II[A,td,\{P12,1,1,0,1,1\},1,\{1,0,0\},\{0,0,0\}]}} && = \int  \frac{\dFtdA\pent{1,2}{11011}}{(k_1 \cdot k_2)(k_1 \cdot n)}, \\
  & \texttt{\small{II[A,td,\{P12,1,1,1,0,1\},1,\{0,0,0\},\{0,0,0\}]}} && = \int  \frac{\dFtdA\pent{1,2}{11101}}{(k_1 \cdot k_2)}, \\
  & \texttt{\small{II[A,td,\{P12,1,1,1,0,1\},1,\{0,0,0\},\{0,0,1\}]}} && = \int  \frac{\dFtdA\pent{1,2}{11101}}{(k_1 \cdot k_2)(k_{12} \cdot \bar{n} )}, \\
  & \texttt{\small{II[A,td,\{P12,1,1,1,0,1\},1,\{1,0,0\},\{0,0,0\}]}} && = \int  \frac{\dFtdA\pent{1,2}{11101}}{(k_1 \cdot k_2)(k_1 \cdot n)}, \\
  & \texttt{\small{II[A,td,\{P12,1,1,1,0,1\},1,\{1,0,0\},\{0,0,1\}]}} && = \int  \frac{\dFtdA\pent{1,2}{11101}}{(k_1 \cdot k_2)(k_1 \cdot n)(k_{12} \cdot \bar{n} )}, \\
  & \texttt{\small{II[A,td,\{P12,1,1,1,1,0\},0,\{0,0,0\},\{0,0,1\}]}} && = \int  \frac{\dFtdA\pent{1,2}{11110}}{(k_{12} \cdot \bar{n} )}, \\
  & \texttt{\small{II[A,td,\{P12,1,1,1,1,0\},0,\{1,0,0\},\{0,0,1\}]}} && = \int  \frac{\dFtdA\pent{1,2}{11110}}{(k_1 \cdot n)(k_{12} \cdot \bar{n} )}, \\
  & \texttt{\small{II[A,td,\{P21,0,1,1,1,1\},0,\{0,0,0\},\{0,0,0\}]}} && = \int \dFtdA\pent{2,1}{01111}, \\
  & \texttt{\small{II[A,td,\{P21,0,1,1,1,1\},0,\{0,0,0\},\{0,0,1\}]}} && = \int  \frac{\dFtdA\pent{2,1}{01111}}{(k_{12} \cdot \bar{n} )}, \\
  & \texttt{\small{II[A,td,\{P21,1,0,1,0,1\},0,\{0,0,0\},\{0,0,1\}]}} && = \int  \frac{\dFtdA\pent{2,1}{10101}}{(k_{12} \cdot \bar{n} )}, \\
  & \texttt{\small{II[A,td,\{P21,1,0,1,0,1\},1,\{0,0,0\},\{0,0,0\}]}} && = \int  \frac{\dFtdA\pent{2,1}{10101}}{(k_1 \cdot k_2)}, \\
  & \texttt{\small{II[A,td,\{P21,1,0,1,1,1\},1,\{0,0,0\},\{0,0,0\}]}} && = \int  \frac{\dFtdA\pent{2,1}{10111}}{(k_1 \cdot k_2)}, \\
  & \texttt{\small{II[A,td,\{P21,1,0,1,1,1\},1,\{0,0,0\},\{0,0,1\}]}} && = \int  \frac{\dFtdA\pent{2,1}{10111}}{(k_1 \cdot k_2)(k_{12} \cdot \bar{n} )}, \\
  & \texttt{\small{II[A,td,\{P21,1,1,1,0,1\},0,\{0,0,0\},\{0,0,0\}]}} && = \int \dFtdA\pent{2,1}{11101}, \\
  & \texttt{\small{II[A,td,\{P21,1,1,1,0,1\},1,\{0,0,0\},\{1,0,0\}]}} && = \int  \frac{\dFtdA\pent{2,1}{11101}}{(k_1 \cdot k_2)(k_1 \cdot \bar{n})}, \\
  & \texttt{\small{II[A,td,\{P21,1,1,1,1,0\},0,\{0,0,0\},\{0,0,1\}]}} && = \int  \frac{\dFtdA\pent{2,1}{11110}}{(k_{12} \cdot \bar{n} )}, \\
  & \texttt{\small{II[A,td,\{P21,1,1,1,1,0\},0,\{0,0,0\},\{1,0,0\}]}} && = \int  \frac{\dFtdA\pent{2,1}{11110}}{(k_1 \cdot \bar{n})}, \\
  & \texttt{\small{II[B,tt,\{P12,0,1,0,1,0\},0,\{0,0,1\},\{0,0,1\}]}} && = \int  \frac{\dFttB\pent{1,2}{01010}}{(k_{12} \cdot n)(k_{12} \cdot \bar{n} )}, \\
  & \texttt{\small{II[B,tt,\{P12,0,1,0,1,0\},0,\{0,0,1\},\{1,0,0\}]}} && = \int  \frac{\dFttB\pent{1,2}{01010}}{(k_{12} \cdot n)(k_1 \cdot \bar{n})}, \\
  & \texttt{\small{II[B,tt,\{P12,0,1,0,1,0\},0,\{0,1,0\},\{1,0,0\}]}} && = \int  \frac{\dFttB\pent{1,2}{01010}}{(k_2 \cdot n)(k_1 \cdot \bar{n})}, \\
  & \texttt{\small{II[B,tt,\{P12,1,0,1,0,1\},1,\{0,0,0\},\{0,0,0\}]}} && = \int  \frac{\dFttB\pent{1,2}{10101}}{(k_1 \cdot k_2)}, \\
  & \texttt{\small{II[B,tt,\{P12,1,1,0,1,1\},1,\{0,0,0\},\{0,0,0\}]}} && = \int  \frac{\dFttB\pent{1,2}{11011}}{(k_1 \cdot k_2)},
                                                                         \label{eqb.49}
  \\
  & \texttt{\small{II[B,tt,\{P12,1,1,0,1,1\},1,\{0,1,0\},\{1,0,0\}]}} && = \int  \frac{\dFttB\pent{1,2}{11011}}{(k_1 \cdot k_2)(k_2 \cdot n)(k_1 \cdot \bar{n})}, \\
  & \texttt{\small{II[B,tt,\{P12,1,1,1,0,1\},1,\{0,0,1\},\{0,0,0\}]}} && = \int  \frac{\dFttB\pent{1,2}{11101}}{(k_1 \cdot k_2)(k_{12} \cdot n)}, \\
  & \texttt{\small{II[B,tt,\{P12,1,1,1,1,0\},0,\{0,0,1\},\{0,0,1\}]}} && = \int  \frac{\dFttB\pent{1,2}{11110}}{(k_{12} \cdot n)(k_{12} \cdot \bar{n} )}, \\
  & \texttt{\small{II[B,tt,\{P21,0,1,1,1,1\},0,\{0,0,1\},\{0,0,0\}]}} && = \int  \frac{\dFttB\pent{2,1}{01111}}{(k_{12} \cdot n)}, \\
  & \texttt{\small{II[B,tt,\{P21,1,0,1,0,1\},1,\{0,0,0\},\{0,0,0\}]}} && = \int  \frac{\dFttB\pent{2,1}{10101}}{(k_1 \cdot k_2)}, \\
  & \texttt{\small{II[B,tt,\{P21,1,0,1,0,1\},1,\{0,0,1\},\{0,0,0\}]}} && = \int  \frac{\dFttB\pent{2,1}{10101}}{(k_1 \cdot k_2)(k_{12} \cdot n)}, \\
  & \texttt{\small{II[B,tt,\{P21,1,0,1,1,1\},1,\{0,1,0\},\{0,0,1\}]}} && = \int  \frac{\dFttB\pent{2,1}{10111}}{(k_1 \cdot k_2)(k_2 \cdot n)(k_{12} \cdot \bar{n} )},
                                                                         \label{eqb.56}
  \\
  & \texttt{\small{II[B,tt,\{P21,1,1,1,0,1\},0,\{0,0,1\},\{0,0,0\}]}} && = \int  \frac{\dFttB\pent{2,1}{11101}}{(k_{12} \cdot n)}, \\
  & \texttt{\small{II[B,tt,\{P21,1,1,1,0,1\},1,\{0,0,0\},\{0,0,0\}]}} && = \int  \frac{\dFttB\pent{2,1}{11101}}{(k_1 \cdot k_2)}, \\
  & \texttt{\small{II[B,tt,\{P21,1,1,1,1,0\},0,\{0,0,1\},\{0,0,1\}]}} && = \int  \frac{\dFttB\pent{2,1}{11110}}{(k_{12} \cdot n)(k_{12} \cdot \bar{n} )}, \\
  & \texttt{\small{II[B,tt,\{P21,1,1,1,1,0\},0,\{0,0,1\},\{1,0,0\}]}} && = \int  \frac{\dFttB\pent{2,1}{11110}}{(k_{12} \cdot n)(k_1 \cdot \bar{n})}, \\
  & \texttt{\small{II[B,tt,\{P21,1,1,1,1,1\},0,\{0,0,0\},\{0,0,0\}]}} && = \int \dFttB\pent{2,1}{11111}.
                                                                         \label{eqb.61}
\end{alignat}

\bibliographystyle{JHEP}
\bibliography{sfRRV}
\end{document}